\documentclass[superscriptaddress,aps,pra,nofootinbib,twocolumn,notitlepage,10pt]{revtex4-1}
\pdfoutput=1
\usepackage{graphicx}
\usepackage{amsmath}
\usepackage{amssymb}
\usepackage{comment}
\usepackage{placeins}
\usepackage[caption=false]{subfig}
\usepackage[colorlinks]{hyperref}
\usepackage{hypcap}

\newcommand{\eq}[1]{Eq.~\hyperref[eq:#1]{(\ref*{eq:#1})}}
\renewcommand{\sec}[1]{\hyperref[sec:#1]{Section~\ref*{sec:#1}}}
\newcommand{\app}[1]{\hyperref[app:#1]{Appendix~\ref*{app:#1}}}
\newcommand{\tab}[1]{\hyperref[tab:#1]{Table~\ref*{tab:#1}}}
\newcommand{\fig}[1]{\hyperref[fig:#1]{Figure~\ref*{fig:#1}}}
\newcommand{\figa}[2]{\hyperref[fig:#1]{Figure~\ref*{fig:#1}#2}}
\newcommand{\figx}[2]{\hyperref[fig:#1]{Figure~\ref*{fig:#1}(#2)}}
\newcommand{\thm}[1]{\hyperref[thm:#1]{Theorem~\ref*{thm:#1}}}
\newcommand{\lem}[1]{\hyperref[lem:#1]{Lemma~\ref*{lem:#1}}}
\newcommand{\cor}[1]{\hyperref[cor:#1]{Corollary~\ref*{cor:#1}}}
\newcommand{\defn}[1]{\hyperref[def:#1]{Definition~\ref*{def:#1}}}
\newcommand{\alg}[1]{\hyperref[alg:#1]{Algorithm~\ref*{alg:#1}}}

\def\bra#1{\mathinner{\langle{#1}|}}
\def\ket#1{\mathinner{|{#1}\rangle}}

\newcommand{\ignore}[1]{}

\newcommand{\be}{\begin{equation}}
\newcommand{\ee}{\end{equation}}
\newcommand{\ba}{\begin{eqnarray}}
\newcommand{\ea}{\end{eqnarray}}

\newcommand{\nn}{\nonumber \\}
\newcommand{\kets}[1]{ |{#1} \rangle}

\newcommand{\rad}{\xi}
\newcommand{\bs}[1]{\boldsymbol{#1}}
\newcommand{\wcomp}{\widetilde{\cal O}(N)}

\input{Qcircuit}
\begin{document}

\title{Exponentially more precise quantum simulation of fermions I:\\ Quantum chemistry in second quantization}

\date{\today}
\author{Ryan Babbush}
\email[Corresponding author: ]{ryanbabbush@gmail.com}
\affiliation{Department of Chemistry and Chemical Biology, Harvard University, Cambridge, MA 02138}
\affiliation{Google, Venice, CA 90291, USA}
\author{Dominic W. Berry}
\email[Corresponding author: ]{dominic.berry@mq.edu.au}
\affiliation{Department of Physics and Astronomy, Macquarie University, Sydney, NSW 2109, Australia}
\author{Ian D. Kivlichan}
\affiliation{Department of Chemistry and Chemical Biology, Harvard University, Cambridge, MA 02138}
\affiliation{Department of Physics, Harvard University, Cambridge, MA 02138, USA}
\author{Annie Y. Wei}
\affiliation{Department of Chemistry and Chemical Biology, Harvard University, Cambridge, MA 02138}
\author{Peter J. Love}
\affiliation{Department of Physics and Astronomy, Tufts University, Medford, MA 02155}
\author{Al\'{a}n Aspuru-Guzik}
\email[Corresponding author: ]{aspuru@chemistry.harvard.edu}
\affiliation{Department of Chemistry and Chemical Biology, Harvard University, Cambridge, MA 02138}

\begin{abstract}
We introduce novel algorithms for the quantum simulation of molecular systems which are asymptotically more efficient than those based on the Trotter-Suzuki decomposition. We present the first application of a recently developed technique for simulating Hamiltonian evolution using a truncated Taylor series to obtain logarithmic scaling with the inverse of the desired precision, an exponential improvement over all prior methods. The two algorithms developed in this work rely on a second quantized encoding of the wavefunction in which the state of an $N$ spin-orbital system is encoded in ${\cal O}(N)$ qubits.
Our first algorithm requires at most $\widetilde{\cal O}(N^8 t)$ gates. Our second algorithm involves on-the-fly computation of molecular integrals, in a way that is exponentially more precise than classical sampling methods, by using the truncated Taylor series simulation technique. Our second algorithm has the lowest gate count of any approach to second quantized quantum chemistry simulation in the literature, scaling as $\widetilde{\cal O}(N^{5} t)$. The approaches presented here are readily applicable to a wide class of fermionic models, many of which are defined by simplified versions of the chemistry Hamiltonian.
\end{abstract}
\maketitle

\section{Introduction}
\label{sec:intro}

As small, fault-tolerant quantum computers come increasingly close to viability \cite{Martinis2014,Martinis2015,Nigg2014,Corcoles2015} there has been substantial renewed interest in quantum simulating chemistry due to the low qubit requirements and industrial importance of the electronic structure problem.
A recent series of papers tried to estimate the resources required to quantum simulate a small but classically intractable molecule \cite{Wecker2014,Hastings2015,Poulin2014,McClean2014,BabbushTrotter}.
Although qubit requirements seem modest, initial predictions of the time required were daunting. Using arbitrarily high-order Trotter formulas,  the tightest known bound on the gate count of the second quantized, Trotter-based quantum simulation of chemistry is $\widetilde{\cal O}(N^8 t / \epsilon^{o(1)})$ \cite{Berry2006,Wiebe2011}\footnote{We use the typical computer science convention that $f\in \Theta(g)$, for any functions $f$ and $g$,
if $f$ is asymptotically upper and lower bounded by multiples of $g$, ${\cal O}$ indicates an asymptotic upper bound, $\widetilde{{\cal O}}$ indicates an asymptotic upper bound up to polylogarithmic factors, $\Omega$ indicates the asymptotic lower bound and $f \in o(g)$ implies $f / g \rightarrow 0$ in the asymptotic limit.}, where $\epsilon$ is the precision required and $N$ is the number of spin-orbitals. However, using significantly more practical Trotter decompositions, the best known gate complexity for this quantum algorithm is $\widetilde{\cal O}(N^9 \sqrt{t^3 / \epsilon})$ \cite{Hastings2015}.

Fortunately, numerics indicated that the average circuit depth for real molecules may be closer to $\widetilde{\cal O}(N^6 \sqrt{t^3 / \epsilon})$ \cite{Poulin2014}, or $\widetilde{\cal O}(Z^3 N^4 \sqrt{t^3 / \epsilon})$ when only trying to simulate ground states, where $Z$ is the largest nuclear charge for the molecule \cite{BabbushTrotter}.
While this improved scaling restores hope that fault-tolerant devices will have an impact on some classically intractable chemistry problems, the Trotter-based quantum simulation of large (e.g. $N > 500$) molecules still seems prohibitively costly \cite{BabbushTrotter,Gibney2014,Mueck2015}.
This limitation would preclude simulations of many important molecular systems, such as those involved in biological nitrogen fixation and high-$T_c$ superconductivity \cite{Gibney2014,Mueck2015}.

The canonical quantum algorithm for quantum chemistry, based on the Trotter-Suzuki decomposition which was first applied for universal quantum simulation in \cite{Lloyd1996,Abrams1997}, was introduced nearly one decade ago \cite{Aspuru-Guzik2005}. This approach was later refined for implementation with a set of universal quantum gates in \cite{Whitfield2010}. With the exception of the adiabatic algorithm described in \cite{BabbushAQChem} and a classical variational optimization strategy making use of a quantum wavefunction ansatz described in \cite{Peruzzo2013}, all prior quantum algorithms for chemistry have been based on Trotterization \cite{Jones2012,Veis2010,Wang2014,Whitfield2013b,Whitfield2015,Li2011,Yung2013,Toloui2013}.

Trotter-Suzuki approaches were also applied to simulation of evolution under sparse Hamiltonians with the entries given by an oracle \cite{Aharonov2003,Berry2007}.
A related problem is the simulation of continuous query algorithms;
in 2009, Cleve \emph{et al.} showed how to achieve such simulation with exponentially fewer discrete queries than Trotterization in terms of $1/\epsilon$ \cite{Cleve2009}.
The algorithm of \cite{Cleve2009} still required a number of ancilla qubits that scaled polynomially in $1/\epsilon$, but this limitation was overcome in \cite{Berry2014} which demonstrated that the ancilla register in \cite{Cleve2009} could be compressed into exponentially fewer qubits.
In \cite{Berry2013,Berry2015}, Berry \emph{et al.}~combined the results of \cite{Berry2007,Aharonov2003,Cleve2009,Berry2014} to show exponentially more precise sparse Hamiltonian simulation techniques.
A major contribution of \cite{Berry2013} was to use oblivious amplitude amplification to make the algorithm from \cite{Cleve2009,Berry2014} deterministic, whereas prior versions had relied on probabilistic measurement of ancilla qubits.
An improvement introduced in \cite{Berry2015} was to show how to simulate arbitrary Hamiltonians using queries that are not self-inverse (a requirement of the procedure in \cite{Berry2013}).
We focus on the methodology of \cite{Berry2015} which is relatively self-contained.

The algorithm of \cite{Berry2015} approximates the propagator using a Taylor series expansion rather than the Trotter-Suzuki decomposition. By dividing the desired evolution into a number of simulation segments proportional to the Hamiltonian norm, one can truncate the Taylor series at an order which scales logarithmically in the inverse of the desired precision \cite{Berry2015}. The truncated Taylor series must be expressed as a weighted sum of unitary operators. To simulate the action of this operator, one first initializes the system along with an ancilla register that indexes all terms in the Taylor series sum. The ancilla register is then put in a superposition state with amplitudes proportional to the coefficients of terms in the Taylor series sum. Next, an operator is applied to the system which coherently executes a single term in the Taylor series sum that is selected according to the ancilla register in superposition. Finally, by applying the transpose of the procedure which prepares the ancilla register, one probabilistically simulates evolution under the propagator. The algorithm is made deterministic using an oblivious amplitude amplification procedure from \cite{Berry2013}.

This is the first paper of a two-paper series which applies the algorithm of \cite{Berry2015} to quantum chemistry simulation. The algorithms discussed in this paper employ a second quantized encoding of the Hamiltonian, where we dynamically perform the Jordan-Wigner transformation \cite{Jordan1928,Somma2002} on the quantum computer. In the second paper of this series, we use a compressed, first quantized encoding of the wavefunction which requires a number of qubits that scales almost linearly with the number of electrons \cite{BabbushSparse2}.

In the present paper we develop two new algorithms for the application of the Hamiltonians terms, which we refer to as the ``database'' algorithm and the ``on-the-fly'' algorithm. In the database algorithm, the ancilla register's superposition state is prepared with amplitudes from a precomputed classical database. In the on-the-fly algorithm, those amplitudes are computed and prepared on-the-fly, in a way that is exponentially more precise than classically possible.

\section{Overview of Results}
\label{sec:overview}

The simulation procedure described in \cite{Berry2015} assumes the ability to represent the Hamiltonian as a weighted sum of unitaries which can be individually applied to a quantum state. Specifically, we must be able to express the simulation Hamiltonian as 
\begin{equation}
H = \sum_{\gamma=1}^{\Gamma} W_\gamma H_\gamma
\label{eq:unit_sum}
\end{equation}
where the $W_\gamma$ are complex-valued scalars\footnote{The convention of \cite{Berry2015} requires that the $W_\gamma$ are real, non-negative scalars. This treatment remains general as arbitrary phases can be factored into the $H_\gamma$. However, we break with that convention and allow the $W_\gamma$ to take arbitrary complex values. This is done for pedagogical purposes: so that we may separately describe computation of the $H_\gamma$ and the $W_\gamma$ for the chemistry Hamiltonian. Consequentially, our \eq{W} differs from the analogous equation in \cite{Berry2015} by a complex conjugate operator.}, the $H_\gamma$ are unitary operators and a mechanism is available for selectively applying the $H_\gamma$. Using the Jordan-Wigner  transformation \cite{Jordan1928,Somma2002} or the Bravyi-Kitaev transformation \cite{Bravyi2002,Seeley2012,Tranter2015}, the second quantized molecular Hamiltonian can be mapped to a sum of $\Gamma \in {\cal O}(N^4)$ local Hamiltonians. Since these local Hamiltonians are each a tensor product of Pauli operators multiplied by some coefficient, they automatically satisfy the form of \eq{unit_sum}.

We will need a circuit referred to in \cite{Berry2015} as $\textsc{select}(H)$ which is queried within the algorithm such that
\begin{equation}
\textsc{select}\left(H\right) \ket{\gamma} \ket{\psi} = \ket{\gamma} H_\gamma\ket{\psi}.
\label{eq:selectH}
\end{equation}
One could construct $\textsc{select}(H)$ by storing all the Pauli strings in a database. However, accessing this data would have time complexity of at least  $\Omega(\Gamma)$. Instead, we compute and apply the Pauli strings using ${\cal O}(N)$ gates (which can be parallelized to ${\cal O}(1)$ circuit depth) by dynamically performing the Jordan-Wigner transformation on the quantum hardware.

The algorithm of \cite{Berry2015} also requires an operator that we refer to as $\textsc{prepare}(W)$ which applies the mapping
\begin{equation}
\label{eq:prepareW}
\textsc{prepare}\left(W\right)\ket{0}^{\otimes \log \Gamma} = \sqrt{\frac{1}{\Lambda}}\sum_{\gamma = 1}^{\Gamma} \sqrt{W_\gamma}\ket{\gamma}
\end{equation}
where 
\begin{equation}
\label{eq:Lambda}
\Lambda \equiv \sum_{\gamma = 1}^{\Gamma} \left | W_\gamma \right |, \quad \quad \Lambda \in {\cal O}\left(N^4\right)
\end{equation}
is a normalization factor that will turn out to have significant ramifications for the algorithm complexity. In the first of two algorithms discussed in this paper, we implement $\textsc{prepare}(W)$ using a database via a sequence of totally controlled rotations at cost ${\cal O}(\Gamma)$. Because our first approach uses a database to store classically precomputed values of $W_\gamma$ in order to implement $\textsc{prepare}(W)$, we refer to the first algorithm as the ``database'' algorithm.

While we suggest a different strategy in \sec{hamiltonian}, a database could also be used to construct $\textsc{select}(H)$.
That is, a controlled operation is performed which applies $H_1$ if $\gamma=1$, followed by a controlled operation which performs $H_2$ if $\gamma=2$, and so forth.
This would result in a slightly higher gate count than $\textsc{prepare}(W)$, because each of the $\Gamma$ controlled operations must act on ${\cal O}(\log N)$ qubits even if the Bravyi-Kitaev transformation is used. Nevertheless, this might represent a simpler solution than our construction of $\textsc{select}(H)$ for early experimental implementations.

Our second algorithm involves modifications to the algorithm of \cite{Berry2015} which allows us to avoid some of this overhead. We exploit the fact that the chemistry Hamiltonian is easy to express as a special case of \eq{unit_sum} in which the coefficients are defined by integrals such as
\begin{equation}
\label{eq:int_w}
W_{\gamma} = \int_{\cal Z} \! w_{\gamma}\left(\vec z\right) \, d\vec z.
\end{equation}
Because our approach involves computing integrals on-the-fly, we refer to the second algorithm as the ``on-the-fly'' algorithm. We begin by numerically approximating the integrals as finite Riemann sums such as
\begin{equation}
\label{eq:int_des}
W_{\gamma} \approx \frac{\cal V}{\mu} \sum_{\rho = 1}^{\mu} w_\gamma \left(\vec z_\rho \right)
\end{equation}
where $\vec z_\rho$ is a point in the integration domain at grid point $\rho$. 
Equation \eqref{eq:int_des} represents a discretization of the integral in \eq{int_w} using $\mu$ grid points where the domain of the integral, denoted as ${\cal Z}$, has been truncated to have total volume ${\cal V}$.
This truncation is possible because the functions $w_\gamma(\vec z)$ can be chosen to decay exponentially for the molecular systems usually studied in chemistry. Note that this might not be true for other systems, such as conducting metals.

Our algorithm is effectively able to numerically compute this integral
with complexity logarithmic in the number of grid points.
It might be thought that this is impossible, because methods of
evaluating numerical integrals on quantum computers normally only give a
square-root speedup over classical Monte-Carlo algorithms \cite{Abrams1999fast}.
The difference here is that we do not output the value of the integral.
The value of the integral is only used to control the weight of a term
in the Hamiltonian under which the state evolves.

We construct a circuit which computes the values of $w_\gamma (\vec z_\rho)$ for the quantum chemistry Hamiltonian with $\wcomp$ gates. We call this circuit $\textsc{sample}(w)$ and define it by its action,
\begin{equation}
\label{eq:samplew}
\textsc{sample}\left(w\right)\ket{\gamma}\ket{\rho}\ket{0}^{\otimes \log M} = \ket{\gamma}\ket{\rho}\ket{\widetilde{w}_{\gamma} \left(\vec z_\rho\right)},
\end{equation}
where $\widetilde{w}_\gamma (\vec z_\rho)$ is the binary representation of $w_\gamma (\vec z_\rho)$ using $\log M$ qubits.

By expanding the $W_\gamma$ in \eq{unit_sum} in terms of the easily
computed $w_\gamma(\vec z)$ as in \eq{int_des}, we are able to compute
analogous amplitudes to those in \eq{prepareW} in an efficient fashion.
Thus, we no longer need the database that characterizes that algorithm.
State preparation where the state coefficients can be computed on
the quantum computer is more efficient than when
they are stored on, and accessed from, a database \cite{Grover2000}.
The worst-case complexity is the square root of the dimension (here it would be ${\cal O}(\sqrt{\Gamma\mu})$), whereas the database
state preparation has complexity linear in the dimension (which is ${\cal O}(\Gamma)$ for $W_\gamma$).
Here this would not be an improvement, as we have increased the
dimension in the discretization of the integral.

However, the worst-case complexity is only if the amplitudes can take
arbitrary values (as this would enable a search algorithm, where the
square root of the dimension is optimal \cite{Grover1996}).
If the amplitudes differ only by phases, then complexity of the state preparation is logarithmic in the dimension.
We therefore decompose each $w_\gamma(\vec z)$ into a sum of terms which
differ only by a sign.
Then, although the dimension is increased, the complexity of the state
preparation is reduced.
The decomposition is of the form
\begin{equation}
\label{eq:unit_decomp}
w_\gamma\left(\vec z\right) \approx \zeta \sum_{m=1}^{M} w_{\gamma,m} \left(\vec z\right), \quad  \quad w_{\gamma, m}\left(\vec z\right) \in \left\{-1,+1\right\},
\end{equation}
where 
\begin{equation}
\zeta \in \Theta\left(\frac{\epsilon}{\Gamma {\cal V} t}\right),\quad \quad M \in \Theta\left(\max_{\vec z,\gamma} \left |w_\gamma \left(\vec z\right)\right | / \zeta\right).
\end{equation}
In turn, we can express the Hamiltonian as a sum of unitaries weighted by identical amplitudes which differ only by an easily computed sign,
\begin{equation}
\label{eq:onthefly}
H = \frac{\zeta {\cal V}}{\mu} \sum_{\gamma = 1}^{\Gamma} \sum_{m=1}^{M} \sum_{\rho = 1}^{\mu} w_{\gamma, m}\left(\vec z_\rho\right) H_\gamma.
\end{equation}

As discussed above, the state preparation needed can be performed much more efficiently because the amplitudes are now identical up to a phase. By making a single query to $\textsc{sample}(w)$ and then performing phase-kickback we can implement the operator $\textsc{prepare}(w)$ whose action is
\begin{equation}
\label{eq:preparew}
\textsc{prepare}\left(w\right)\ket{0}^{\otimes \log \left(L\right)} = \sqrt{\frac{1}{\lambda}} \sum_{\ell = 1}^{L} \sqrt{\frac{\zeta {\cal V}}{\mu} w_{\gamma, m}\left(\vec z_\rho\right) }  \ket{\ell}
\end{equation}
where $\ket{\ell} = \ket{\gamma}\ket{m} \ket{\rho}$, $L \in \Theta(\Gamma M \mu)$ and
\begin{equation}
\label{eq:lambda}
\lambda = L \frac{\zeta {\cal V}}{\mu} \in \Theta\left(\Gamma {\cal V} \max_{\vec z,\gamma}\left |w_\gamma \left(\vec z\right)\right | \right)
\end{equation}
is a normalization factor that will turn out to have significant ramifications for the algorithm complexity. Later, we will show that $\lambda \in \widetilde{\cal O}(N^4)$ and that $\textsc{prepare}(w)$ can be implemented with $\wcomp$ gate count, the cost of a single query to $\textsc{sample}(w)$.

The database algorithm performs evolution under $H$ for time $t$ by making $\widetilde{\cal O}(\Lambda t)$ queries to both $\textsc{select}(H)$ and $\textsc{prepare}(W)$. Because $\textsc{prepare}(W)$ requires ${\cal O}(\Gamma)$ = ${\cal O}(N^4)$ gates, the overall gate count of this approach scales as $\widetilde{\cal O}(N^4 \Lambda t)$. To avoid the overhead from $\textsc{prepare}(W)$, our on-the-fly algorithm exploits a modified version of the truncated Taylor series algorithm which allows for the same evolution by making $\widetilde{\cal O}(\lambda t)$ queries to $\textsc{select}(H)$ and $\textsc{prepare}(w)$. As $\textsc{prepare}(w)$ requires $\wcomp$ gates, the gate count for our on-the-fly algorithm scales as $\widetilde{\cal O}(N \lambda t)$.

The paper is outlined as follows. In \sec{hamiltonian} we introduce the second quantized encoding of the wavefunction and construct $\textsc{select}(H)$. In \sec{simulation} we review the procedure in \cite{Berry2015} to demonstrate our database algorithm which uses $\textsc{select}(H)$ and $\textsc{prepare}(W)$ to perform a quantum simulation which is exponentially more precise than Trotterization. In \sec{integrals} we show that one can modify the procedure in \cite{Berry2015} to allow for essentially the same result while simultaneously computing the integrals on-the-fly, and show how to implement $\textsc{prepare}(w)$ so as to compute the integrals on-the-fly. In \sec{integrands} we bound the errors on the integrals by analyzing the integrands. In \sec{conclusion} we discuss applications of these results and future research directions.

\section{The Hamiltonian Oracle}
\label{sec:hamiltonian}

The molecular electronic structure Hamiltonian describes electrons interacting in a fixed nuclear potential. Using atomic units in which the electron mass, electron charge, Coulomb's constant and $\hbar$ are unity we may write the electronic Hamiltonian as
\begin{align}
\label{eq:electronic}
 H = - \sum_i \frac{\nabla_{\vec r_i}^2}{2} - \sum_{i,j} \frac{Z_i}{|\vec R_i - \vec r_j|} + \sum_{i, j>i} \frac{1}{|\vec r_i - \vec r_j|},
\end{align}
where $\vec R_i$ are the nuclei coordinates, $Z_i$ are the nuclear charges, and $\vec r_i$ are the electron coordinates \cite{Helgaker2002}. We represent the system in a basis of $N$ single-particle spin-orbital functions usually obtained as the solution to a classical mean-field treatment such as Hartree-Fock \cite{Helgaker2002}. Throughout this paper, $\varphi_i(\vec r_j)$ denotes the $i^\textrm{th}$ spin-orbital occupied by the $j^\textrm{th}$ electron which is parameterized in terms of spatial degrees of freedom $\vec r_j$.

In second quantization, antisymmetry is enforced by the operators whereas in first quantization antisymmetry is explicitly in the wavefunction. The second quantized representation of \eq{electronic} is
\begin{equation}
H = \sum_{ij} h_{ij} a_i^{\dagger}a_j + \frac{1}{2} \sum_{ijk\ell} h_{ijk\ell} a_i^{\dagger} a_j^{\dagger} a_k a_\ell
\label{eq:2nd}
\end{equation}
where the one-electron and two-electron integrals are
\begin{align}
 &h_{ij} = \int \varphi_i^*(\vec r) \left(-\frac{\nabla^2}{2} - \sum_{q} \frac{Z_q}{|\vec R_q - \vec r|} \right)\varphi_j(\vec r) \label{eq:single_int}  \,d\vec r  ,  \\
 &h_{ijk\ell} = \int  \frac{ \varphi_i^*(\vec r_1) \varphi_j^*(\vec r_2)  \varphi_\ell(\vec r_1) \varphi_k(\vec r_2) }{|\vec r_1 - \vec r_2|} \, d\vec r_1\, d\vec r_2. \label{eq:double_int}
\end{align}
The operators $a_i^\dagger$ and $a_j$ in \eq{2nd} obey the fermionic anti-commutation relations,
\begin{align}
\label{eq:anticomm}
 \{a_i^\dagger, a_j\} = \delta_{ij}, \quad \quad \{a_i^\dagger, a_j^\dagger\} = \{a_i, a_j\} = 0.
\end{align}
In general, the Hamiltonian in \eq{2nd} contains ${\cal O}(N^4)$ terms, except in certain limits of very large molecules where use of a local basis and truncation of terms lead to scaling on the order of $\widetilde{\cal O}(N^2)$ \cite{McClean2014}. The spatial encoding of \eq{2nd} requires $\Theta(N)$ qubits, one for each spin-orbital.

While fermions are antisymmetric, indistinguishable particles, qubits are distinguishable and have no special symmetries. Accordingly, in order to construct the operator $\textsc{select}(H)$, which applies terms in the second quantized Hamiltonian to qubits as in \eq{selectH}, we will need a mechanism for mapping the fermionic raising and lowering operators in \eq{2nd} to operators which act on qubits. Operators which raise or lower the state of a qubit are trivial to represent using Pauli matrices,
\begin{align}
\label{eq:qubitraise}
\sigma_j^{+} & = \ket{1}\bra{0} = \frac{1}{2}\left(\sigma^{x}_j - i \,\sigma^{y}_j\right) , \\
\sigma_j^{-} & = \ket{0}\bra{1} = \frac{1}{2}\left(\sigma^{x}_j + i\, \sigma^{y}_j\right).
\end{align}
Throughout this paper, $\sigma^x_j$, $\sigma^y_j$ and $\sigma^z_j$ denote Pauli matrices acting on the $j^\textrm{th}$ tensor factor. However, these qubit raising and lowering operators do not satisfy the fermionic anti-commutation relations in \eq{anticomm}. To enforce this requirement we can apply either the  Jordan-Wigner transformation \cite{Jordan1928,Somma2002} or the Bravyi-Kitaev transformation \cite{Bravyi2002,Seeley2012,Tranter2015}.

The action of $a^\dagger_j$ or $a_j$ must also introduce a phase to the wavefunction which depends on the parity (i.e.\ sum modulo 2) of the occupancies of all orbitals with index less than $j$ \cite{Seeley2012}. If $f_j \in \{0,1\}$ denotes the occupancy of orbital $j$ then
\begin{align}
\label{eq:phases}
a^\dagger_j & \ket{f_N \cdots f_{j+1} \,0 \,f_{j-1} \cdots f_1}\nonumber\\
& = \left(-1\right)^{\sum_{s = 1}^{j-1} f_s} \ket{f_N \cdots f_{j+1} \,1\, f_{j-1} \cdots f_1}\\
a_j & \ket{f_N \cdots f_{j+1}\, 1\, f_{j-1} \cdots f_1}\nonumber\\
& = \left(-1\right)^{\sum_{s = 1}^{j-1} f_s} \ket{f_N \cdots f_{j+1} \,0\, f_{j-1} \cdots f_1}\\
a^\dagger_j & \ket{f_N \cdots f_{j+1} \,1\, f_{j-1} \cdots f_1} = 0\\
a_j & \ket{f_N \cdots f_{j+1} \,0\, f_{j-1} \cdots f_1} = 0.
\end{align}
In general, two pieces of information are needed in order to make sure the qubit encoding of the fermionic state picks up the correct phase: the occupancy of the state and the parity of the occupancy numbers up to $j$. The Jordan-Wigner transformation maps the occupancy of spin-orbital $j$ directly into the state of qubit $j$. Thus, in the Jordan-Wigner transformation, occupancy information is stored locally. However, in order to measure the parity of the state in this representation, one needs to measure the occupancies of all orbitals less than $j$. Because of this, the Jordan-Wigner transformed operators are $N$-local, which means that some of the Jordan-Wigner transformed operators are tensor products of up to $N$ Pauli operators. The Jordan-Wigner transformed operators are
\begin{align}
\label{eq:jwup}
a^{\dagger}_j & \equiv \sigma^{+}_j \bigotimes_{s=1}^{j-1} \sigma_s^{z} = \frac{1}{2}\left(\sigma^{x}_j - i \,\sigma^{y}_j\right) \otimes \sigma_{j-1}^z \cdots \otimes \sigma_1^z\\
\label{eq:jwdown}
a_j & \equiv \sigma^{-}_j \bigotimes_{s=1}^{j-1} \sigma_s^{z} = \frac{1}{2}\left(\sigma^{x}_j + i \,\sigma^{y}_j\right) \otimes \sigma_{j-1}^z \cdots \otimes \sigma_1^z.
\end{align}

It would be convenient if we could construct $\textsc{select}(H)$ by applying the Jordan-Wigner transform and acting on the quantum state, one spin-orbital index at a time. For instance, $\textsc{select}(H)$ might control the application of a fermionic operator as follows
\begin{align}
\label{eq:mess}
\ket{ijk\ell}\ket{\psi} & \mapsto  \ket{ijk\ell}a_\ell\ket{\psi}\nonumber\\
& \mapsto  \ket{ijk\ell}a_k a_\ell\ket{\psi}\nonumber\\
& \mapsto  \ket{ijk\ell}a^\dagger_j a_k a_\ell\ket{\psi} \nonumber\\
& \mapsto  \ket{ijk\ell}a_i^\dagger a_j^\dagger a_k a_\ell\ket{\psi}.
 \end{align}
However, the operators $a_j^\dagger$ and $a_j$ are not unitary because the operators $\sigma^+$ and $\sigma^-$ are not unitary. To correct this problem, we add four qubits to the selection register where each of the four qubits indicates whether the $\sigma^x$ or the $\pm i \,\sigma^y$ part of the $\sigma^+$ and $\sigma^-$ operators should be applied for each of the four fermionic operators in a string such as $a^\dagger_i a^\dagger_j a_k a_\ell$. For ease of exposition, we define new fermionic operators which are unitary, $a^\dagger_{j,q}$ and $a_{j,q}$ where $q \in \{0,1\}$,
\begin{align}
\label{eq:jw12up}
a^{\dagger}_{j,0} \equiv \sigma^x_j \bigotimes_{s=1}^{j-1} \sigma_s^{z}, \quad & \quad a^{\dagger}_{j,1} \equiv -i \,\sigma^y_j \bigotimes_{s=1}^{j-1} \sigma_s^{z}, \\
\label{eq:jw12down}
a_{j,0} \equiv \sigma^x_j \bigotimes_{s=1}^{j-1} \sigma_s^{z}, \quad & \quad a_{j,1} \equiv i\,\sigma^y_j \bigotimes_{s=1}^{j-1} \sigma_s^{z}.
\end{align}
We use these definitions to rewrite the Hamiltonian in \eq{2nd} so that it is explicitly a weighted sum of unitary Pauli products of the form we require in \eq{unit_sum},
\begin{align}
\label{eq:3rd}
H & =\sum_{q_1 q_2} \sum_{ij} \frac{h_{ij}}{4} a_{i,q_1}^{\dagger} a_{j,q_2}\nonumber\\
& \quad + \sum_{q_1 q_2 q_3 q_4} \,\sum_{ijk\ell} \frac{h_{ijk\ell}}{32} a_{i,q_1}^{\dagger}  a_{j,q_2}^{\dagger} a_{k,q_3} a_{\ell,q_4}.
\end{align}
Inspection reveals that applying the transformations in \eq{jw12up} and \eq{jw12down} to \eq{3rd} gives the same expression as applying the transformations in \eq{jwup} and \eq{jwdown} to \eq{2nd}. By removing factors of $1/2$ from both transformation operators and instead placing them in \eq{3rd}, we obtain transformation operators that are always unitary tensor products of Pauli operators.

Accordingly, we can implement $\textsc{select}(H)$ in the spirit of \eq{mess} by using four additional qubits and the transformation operators in \eq{jw12up} and \eq{jw12down} so that
\begin{align}
\label{eq:mess2}
\ket{ijk\ell}& \ket{q_1 q_2 q_3 q_4}\ket{\psi} \mapsto  \ket{ijk\ell}\ket{ q_1 q_2 q_3 q_4}a_{\ell,q_4} \ket{\psi}\nonumber\\
& \mapsto  \ket{ijk\ell} \ket{q_1 q_2 q_3 q_4}a_{k,q_3} a_{\ell, q_4} \ket{\psi}\nonumber\\
& \mapsto  \ket{ijk\ell} \ket{q_1 q_2 q_3 q_4}a^\dagger_{j, q_2} a_{k,q_3} a_{\ell, q_4} \ket{\psi}\nonumber\\
& \mapsto  \ket{ijk\ell} \ket{q_1 q_2 q_3 q_4}a^\dagger_{i, q_1} a^\dagger_{j, q_2} a_{k,q_3} a_{\ell, q_4} \ket{\psi}.
 \end{align}
A circuit which implements these operators controlled on the selection register is straightforward to construct. Furthermore, the transformation of the terms can be accomplished in ${\cal O}(1)$ time. Because the Jordan-Wigner transformation is $N$-local, the number of gates required to actually apply the unitaries in $\textsc{select}(H)$ is ${\cal O}(N)$. However, the terms in \eq{jw12up} and \eq{jw12down} are trivial to apply in parallel so that each query takes ${\cal O}(1)$ time.

Whereas the Jordan-Wigner transformation stores occupancy information locally and parity information $N$-locally, the Bravyi-Kitaev transformation stores both parity and occupancy information in a number of qubits that scales as ${\cal O}(\log N)$ \cite{Bravyi2002,Seeley2012,Tranter2015}. For this reason, the operators obtained using the Bravyi-Kitaev basis act on at most ${\cal O}(\log N)$ qubits. It might be possible to apply the Bravyi-Kitaev transformation with ${\cal O}(\log N)$ gates, which would allow for an implementation of $\textsc{select}(H)$ with ${\cal O}(\log N)$ instead of ${\cal O}(N)$ gates. However, the Bravyi-Kitaev transformation is much more complicated and this would not change the asymptotic scaling of our complete algorithm. The reason for this is because the total cost will depend on the sum of the gate count of $\textsc{select}(H)$ and the gate count of $\textsc{prepare}(W)$ or $\textsc{prepare}(w)$, and the latter procedures always require at least ${\cal O}(N)$ gates.

\section{Simulating Hamiltonian Evolution}
\label{sec:simulation}

\begin{table*}[bt]
\caption{Database algorithm parameters and bounds}
\label{tab:parameters}
\begin{tabular}{  c   |   c   |   c }
\hline\hline
Parameter & Explanation & Bound\\
\hline
$\Lambda$ & normalization factor, \eq{Lambda} & ${\cal O}\left(N^4\right)$\\
$r$ & number of time segments, \eq{r} & $\Lambda t/ \ln(2)$\\
$K$ & truncation point for Taylor series, \eq{K} & ${\cal O}\left(\frac{\log\left(r/\epsilon\right)}{\log\log\left(r/\epsilon\right)}\right)$\\
$\Gamma$ & number of terms in unitary decomposition, \eq{unit_sum} & ${\cal O}\left(N^4\right)$\\
$J$ & number of ancilla qubits in selection register, \eq{J} & $\Theta\left(K \log \Gamma\right)$\\
\hline
\end{tabular}
\end{table*}
\begin{table*}[bt]
\caption{Database algorithm operators and gate counts}
\label{tab:operators}
\begin{tabular}{  c   |   c   |   c }
\hline\hline
Operator & Purpose & Gate Count \\
\hline
$\textsc{select}\left(H\right)$ & applies specified terms from decomposition, \eq{selectH} & ${\cal O}\left(N\right)$\\
$\textsc{select}\left(V\right)$ & applies specified strings of terms, \eq{selectV} & $ {\cal O}\left(N K\right) $\\
$\textsc{prepare}\left(W\right)$ & prepares a superposition of states weighted by coefficients, \eq{prepareW} & $ {\cal O}\left(\Gamma\right) $\\
$\textsc{prepare}\left(\beta\right)$ & prepares a superposition of states weighted by coefficients, \eq{prepareB} & $ {\cal O}\left(K \Gamma \right) $\\
${\cal W}$ & probabilistically performs simulation under $H$ for time $t/r$, \eq{W} & $ {\cal O}\left(K \Gamma\right) $\\
$P$ & projects system onto $\ket{0}^{\otimes J}$ state of selection register, \eq{P} & $ \Theta\left(K\log\Gamma\right) $\\
$G$ & amplification operator to implement sum of unitaries, \eq{G} & $ {\cal O}\left(K \Gamma\right) $\\
$\left(PG\right)^r$ & entire algorithm & $ {\cal O}\left(r K \Gamma \right) $\\
\hline
\end{tabular}
\end{table*}

% Start section
Using the method of \cite{Berry2015}, Hamiltonian evolution can be simulated with an exponential improvement in precision over Trotter-based methods by approximating the truncated Taylor series of the time evolution operator $U = e^{- i H t}$. We begin by partitioning the total simulation time $t$ into $r$ segments of $t/r$. For each of these $r$ segments we perform a Taylor expansion of the propagator and truncate the series at order $K$, i.e.
\begin{align}
\label{eq:Ur}
& U_r \equiv e^{-i H t / r}  \approx\sum_{k=0}^K\frac{\left(-iHt/r\right)^k}{k!}\nonumber\\
& =\sum_{k=0}^K \sum_{\gamma_1,\cdots,\gamma_k=1}^\Gamma\frac{\left(-it/r\right)^k}{k!} W_{\gamma_1} \cdots W_{\gamma_k} H_{\gamma_1}\cdots H_{\gamma_k},
\end{align}
where in the second line we have expanded $H$ as in \eq{unit_sum}. Notice that if we truncate the series at order $K$, we incur error
\begin{equation}
\label{eq:error}
{\cal O}\left(\frac{\left(\left\| H \right\| t / r\right)^{K + 1}}{\left(K+1\right)!}\right).
\end{equation}
If we wish for the total simulation to have error less than $\epsilon$, each segment must have error less than $\epsilon/r$. Accordingly, if we set $r \geq \| H \| t$ then our total simulation will have error at most $\epsilon$ if
\begin{equation}
\label{eq:K}
K \in {\cal O}\left(\frac{\log\left(r/\epsilon\right)}{\log\log\left(r/\epsilon\right)}\right).
\end{equation}

We now discuss how one can actually implement the truncated evolution operator in \eq{Ur}. First note that the sum in \eq{Ur} takes the form
\begin{align}
\label{eq:bV}
\widetilde{U} =\sum_{j} \beta_j V_j, \quad \quad \quad & \quad j \equiv \left(k, \gamma_1, \cdots, \gamma_k\right) , \nonumber\\
\beta_j \equiv \frac{t^k}{r^k k!} W_{\gamma_1} \cdots W_{\gamma_k}, \quad &  V_j \equiv \left(-i\right)^k H_{\gamma_1} \cdots H_{\gamma_k} , 
\end{align}
where the $V_j$ are unitary and $\widetilde{U}$ is close to unitary. Our simulation uses an ancillary ``selection'' register $\ket{j} = \ket{k}\ket{\gamma_1} \cdots \ket{\gamma_K}$ where $0\leq k\leq K$ and $1\leq \gamma_\upsilon\leq \Gamma$ for all $\upsilon$. We will encode $k$ in unary, which requires $\Theta(K)$ qubits, so that $\ket{k}=\ket{1^k0^{K-k}}$. Additionally, we encode each $\ket{\gamma_\upsilon}$ in binary using $\Theta(\log \Gamma)$ qubits. While we need $K$ of the $\ket{\gamma_\upsilon}$ registers, we note that only $k$ will actually be in use for a given value of $\ket{k}$. The total number of ancilla qubits required for the selection register $\ket{j}$, denoted as $J$, scales as
\begin{equation}
\label{eq:J}
J \in \Theta\left(K\log \Gamma\right) = {\cal O}\left(\frac{\log\left(N\right)\log\left(r/\epsilon\right)}{\log\log\left(r/\epsilon\right)}\right).
\end{equation}

% Select V
By making ${\cal O}(K)$ queries to $\textsc{select}(H)$ from \sec{simulation}, we can implement an operator to apply the $V_j$ which is referred to in \cite{Berry2015} as $\textsc{select}(V)$,
\begin{equation}
\label{eq:selectV}
\textsc{select}\left(V\right)\ket{j}\ket{\psi} =\ket{j}V_j\ket{\psi},
\end{equation}
where the $V_j$ are defined as in \eq{bV}. This is equivalent to $k$ applications of $\textsc{select}(H)$, using each of the $\ket{\gamma_\upsilon}$ registers, together with $k$ multiplications by $-i$.
In order to obtain $k$ applications of $\textsc{select}(H)$, we may perform a controlled form of $\textsc{select}(H)$ $K$ times, with each successive qubit in the unary representation of $k$ as the control.
Given that the gate count for $\textsc{select}(H)$ scales as ${\cal O}(N)$, we can implement $\textsc{select}(V)$ with ${\cal O}(N K)$ gates. Applying the Pauli strings in parallel leads to circuit depths of ${\cal O}(1)$ and ${\cal O}(K)$, respectively.\tab{parameters} lists relevant parameters along with their bounds in our database algorithm. \tab{operators} lists relevant operators and their gate counts in our database algorithm.

% Prepare \beta
We will also need an operator that we refer to as $\textsc{prepare}(\beta)$, which initializes a state,
\begin{equation}
\label{eq:prepareB}
\textsc{prepare}\left(\beta\right)\ket{0}^{\otimes J} = \sqrt{\frac{1}{s}}\sum_{j} \sqrt{\beta_j}\ket{j},
\end{equation}
where $s$ is a normalization factor. To implement $\textsc{prepare}(\beta)$ we first prepare the state
\begin{equation}
\left( \sum_{k=0}^K \frac{(\Lambda t/r)^k}{k!} \right)^{-1/2} \sum_{k=0}^K \sqrt{\frac{(\Lambda t/r)^k}{k!}}\ket{k}.
\end{equation}
Using the convention that $R_y(\theta) \equiv \exp [-i\, \theta\, \sigma^y / 2 ]$, we apply $R_y (\theta_1)$ to the first qubit of the unary encoding for $k$ followed by $R_y(\theta_k)$ to the $k$th qubit controlled on qubit $k-1$ for all $k \in [2, K]$ sequentially, where
\begin{equation}
\label{eq:theta_k}
\theta_k \equiv 2\arcsin{\left(\sqrt{ 1- \frac{\left(\Lambda t/r\right)^{k-1}}{\left(k-1\right)!} \left(\sum_{q = k}^K \frac{\left(\Lambda t/r\right)^{q}}{q!}\right)^{-1}} \right)}.
\end{equation}
To each of the $K$ remaining components of the selection register $\ket{\gamma_1} \cdots \ket{\gamma_K}$, we apply $\textsc{prepare}(W)$ once.
\begin{figure}[t]
\[\Qcircuit @R 1em @C 0.25em {
	&&&&&\lstick{\kets{0}} &\gate{R_y\left(\theta_1\right)} & \ctrl{1} & \qw &&&& \cdots &&&& & \qw & \qw & \qw \\
	&&&&&\lstick{\kets{0}} &\qw & \gate{R_y\left(\theta_2\right)} & \qw &&&& \cdots &&&& & \qw & \qw & \qw \\
	&\vdots &&&&&&&&&&& \ddots&& &&  & \vdots & \\\\
	&&&&&\lstick{\kets{0}} &\qw & \qw & \qw &&&& \cdots& &&& & \ctrl{1} & \qw & \qw \\
	&&&&&\lstick{\kets{0}} &\qw & \qw & \qw &&&& \cdots&&& && \gate{R_y\left(\theta_K\right)} & \qw & \qw \\
&&&&&\lstick{\kets{0}^{\otimes \log\Gamma}} & \qw & \gate{\textsc{prepare}\left(W\right)}  &\qw & \qw& \qw & \qw &\qw & \qw& \qw & \qw &\qw & \qw& \qw & \qw  \\
\vdots &&&&&&&  \vdots&&& &&&&&& & &&&&&&& \\\\
&&&&&\lstick{\kets{0}^{\otimes \log\Gamma}} & \qw & \gate{\textsc{prepare}\left(W\right)}  &\qw & \qw & \qw & \qw &\qw & \qw& \qw & \qw&\qw & \qw& \qw & \qw \\
}\]
\caption{\label{fig:B_circuit} The circuit for $\textsc{prepare}(\beta)$ as described in \eq{prepareB}. An expression for $\theta_k$ is given in \eq{theta_k}. $\textsc{prepare}(W)$ is implemented using a precomputed classical database.}
\end{figure}
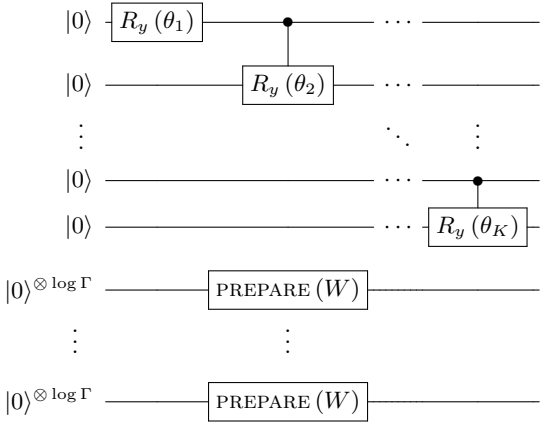
In principle, we only need to perform $\textsc{prepare}(W)$ $k$ times, because the registers past $k$ are not used.
However, it is simpler to perform $\textsc{prepare}(W)$ $K$ times, because it does not require control on $k$.

Using results from \cite{Shende2006}, $\textsc{prepare}(W)$ can be implemented with ${\cal O}(\Gamma)$ gates by using a classically precomputed database of the $\Gamma$ molecular integrals. The gate count for $\textsc{prepare}(\beta)$ thus scales as ${\cal O}(K \Gamma)$. However, this construction is naturally parallelized to depth ${\cal O}(K) + {\cal O}(\Gamma)$. A circuit implementing $\textsc{prepare}(\beta)$ is shown in \fig{B_circuit}.

% general
The general strategy for implementing the truncated evolution operator in \eq{bV} becomes clear if we consider what happens to state $\ket{\psi}$ when we apply $\textsc{prepare}(\beta)$ followed by the operator $\textsc{select}(V)$,
\begin{equation}
\label{eq:BV}
\textsc{select}\left(V\right) \textsc{prepare}\left(\beta\right) \ket{0}^{\otimes J}\! \ket{\psi} =\sum_{j}  \sqrt{\frac{\beta_j}{s}} \ket{j} V_j \ket{\psi}.
\end{equation}
The similarity of this state to the state $\widetilde{U}\ket{\psi}$ motivates the operator
\begin{align}
\label{eq:W}
& {\cal W} \equiv \left( \textsc{prepare}\left(\beta\right)\otimes \openone\right)^\top \! \textsc{select}\left(V\right)\left(\textsc{prepare}\left(\beta\right) \otimes \openone\right)\nonumber\\
& {\cal W} \ket{0}^{\otimes J}\ket{\psi} = \frac{1}{s}\ket{0}\widetilde{U}\ket{\psi} + \sqrt{1 - \frac{1}{s^2}}\ket{\Phi},
\end{align}
where $\ket{\Phi}$ is a state with the ancilla qubits orthogonal to $\ket{0}^{\otimes J}$. Note that in \cite{Berry2015}, the authors use the convention that all $W_\gamma$ are positive and phases are incorporated into the operators $H_\gamma$. Since we depart from that convention for reasons described in \sec{overview}, the second application of $\textsc{prepare}(\beta)$ in \eq{W} is the transpose of the first application, in contrast to \cite{Berry2015} where the conjugate transpose is used instead.

At this point, we choose the number of segments to be
\begin{equation}
\label{eq:r}
r = \Lambda t / \ln\left(2\right).
\end{equation}
Since $\Lambda \geq \| H \|$, our choice of $K$ in \eq{K} remains valid. The additional factor of $1 / \ln(2)$ is included to satisfy a requirement of oblivious amplitude amplification as described in \cite{Berry2013} so that
\begin{equation}
\label{eq:s}
s = \sum_{j} \left | \beta_j  \right |= \sum_{k=0}^{K} \frac{1}{k!} \ln\left(2\right)^k \approx 2.
\end{equation}
We now define a projection operator $P$ onto the target state, which has support on the empty ancilla register,
\begin{align}
\label{eq:P}
& P \equiv \left(\ket{0}\!\bra{0}\right)^{\otimes J} \! \otimes \openone ,\nonumber \\
& P {\cal W} \ket{0}^{\otimes J} \!\ket{\psi} = \frac{1}{s}\ket{0}^{\otimes J} \! \widetilde{U} \ket{\psi}.
\end{align}
We also define the amplification operator,
\begin{equation}
\label{eq:G}
G \equiv - {\cal W} R {\cal W}^\dagger R {\cal W},
\end{equation}
where $R = \openone - 2 P$ is the reflection operator. With these definitions, we follow the procedure in \cite{Berry2015} which uses the oblivious amplitude amplification procedure of \cite{Berry2013} to deterministically execute the intended unitary. We use $G$ in conjunction with $P$ to amplify the target state,
\begin{equation}
PG\ket{0}\ket{\psi}=\ket{0}\left(\frac{3}{s}\widetilde{U}-\frac{4}{s^3}\widetilde{U}\widetilde{U}^\dagger\tilde{U}\right)\ket{\psi}.
\end{equation}
Recalling the definition of $U_r$ in \eq{Ur}, our choice of $K$ in \eq{K} and $r=\Lambda t/\ln 2$ imply that
\begin{equation}
\left\|PG\ket{0}\ket{\psi}-\ket{0}U_r\ket{\psi}\right\| \in {\cal O}\left(\epsilon/r\right),
\end{equation}
so that the total error from applying oblivious amplitude amplification to all the segments will again be order $\epsilon$.

% Complexity
The gate count of the entire algorithm is thus $r$ times the cost of implementing $\textsc{select}(V)$ plus the cost of implementing $\textsc{prepare}(\beta)$. Though we implement $\textsc{select}(V)$ with ${\cal O}(N K)$ gates, our brute-force construction of $\textsc{prepare}(W)$ led to a gate count for $\textsc{prepare}(\beta)$ which scales as ${\cal O}(K \Gamma)$. Thus, the total gate count of our database algorithm scales as
\begin{equation}
\label{eq:database_complex1}
{\cal O}\left(r K \Gamma\right) = {\cal O}\left(\frac{N^4 \Lambda t \log \left(N t  / \epsilon\right)}{\log \log \left(N t / \epsilon\right)}\right) =  \widetilde{\cal O}\left(N^8 t\right).
\end{equation}
While this bound suggests an exponentially more precise algorithm than those based on Trotterization, in the remainder of our paper we discuss an even more efficient algorithm with improved dependence on $N$.

\section{Evolution Under Integral Hamiltonians}
\label{sec:integrals}

In \sec{simulation} we analyzed the database algorithm for quantum simulating chemistry Hamiltonians in a manner that is exponentially more precise than Trotterization. The most costly part of that procedure is the implementation of $\textsc{prepare}(W)$ as in \eq{prepareW}, which prepares a superposition state with amplitudes that are given by integrals over spin-orbitals as in \eq{single_int} and \eq{double_int}. Instead of classically precomputing these integrals and implementing $\textsc{prepare}(W)$ with a database, the strategy we introduce is to numerically sample the integrals on-the-fly using the quantum computer. Because of this, we call this the ``on-the-fly'' algorithm. To accomplish this, we discretize the integrals as sums and design a circuit which returns the integrand of these integrals at particular domain points. The motivation for approximating integrals as sums comes from a direct analogy between the discretization of time in the Taylor series approach for simulating time-dependent Hamiltonians \cite{Berry2015} and the discretization of space in Riemann integration.

In \cite{Berry2015}, the time-ordered exponential is approximated by a Taylor series up to order $K$, and the integrals are then discretized as follows on each segment,
\begin{align}
& {\cal T} \exp\left[-i \int_{0}^{t/r} \!\! H\left(t\right) d t \right] \nn
&\approx \sum_{k=0}^K \frac{(-i)^k}{k!}\int_{0}^{t/r} {\cal T} H\left(t_k\right)\dots H\left(t_1\right) d\bs t \nn
&\approx \sum_{k=0}^K \frac{(-it/r)^k}{\mu^k k!}\sum_{j_1,\ldots,j_k=0}^{\mu-1} H\left(t_{j_k}\right)\dots H\left(t_{j_1}\right).
\end{align}
Now let us suppose that our Hamiltonian does not change in time, but instead that the Hamiltonian itself is given as a definite integral over the spatial region ${\cal Z}$ so that
\begin{equation}
\label{eq:int_H}
H = \int_{\cal Z} {\cal H}\left(\vec z\right) d\vec z.
\end{equation}
The second quantized Hamiltonian given by \eq{3rd} is similar to this, except it includes both terms $h_{ij}$, which are integrals over one spatial coordinate, and $h_{ijk\ell}$, which are integrals over two spatial coordinates. While those integrands are also defined over all space, the integrands decay exponentially so we can approximate them as definite integrals over the finite region ${\cal Z}$, having volume ${\cal V}$. Then we can approximate the integral by
\begin{align}
\label{eq:double}
H & \approx \int_{\cal Z} {\cal H}\left(\vec z\right) d \vec z \approx \frac{\cal V}{\mu} \sum_{\rho=1}^{\mu} {\cal H}\left(\vec z_\rho\right),
\end{align}
where $\vec z_\rho$ is a point in the domain ${\cal Z}$ at the $\rho^{\textrm{th}}$ grid point.

% Start integral simulation
As in \sec{simulation}, we begin by dividing $t$ into $r$ segments. We turn our attention to a single segment,
\begin{align}
\label{eq:Ur1}
U_r & \approx \exp\left[-i \frac{t}{r} \int_{\cal Z}  {\cal H}\left(\vec z\right) d\vec z \right]\nonumber \\
& \approx \sum_{k=0}^K\frac{\left(-it/r\right)^k}{k!}\left(\int_{\cal Z} {\cal H}\left(\vec z\right)\, d\vec z\right)^k \nonumber\\
& = \sum_{k=0}^K\frac{\left(-it/r\right)^k}{k!}\int_{\cal Z} {\cal H}\left(\vec z_1\right)\cdots {\cal H}\left(\vec z_k\right)\, d\bs{\vec z}
\end{align}
where in the second line we have performed a Taylor expansion of the propagator and truncated at order $K$. In \eq{Ur1}, the bolded symbol $\bs{\vec z}$ indicates a vector of vectors. Like before, if $r \geq \|H \| t$ then the relationship between $K$ and $\epsilon$ is given by \eq{K}. To approximate the integral, we divide it into $\mu$ regions of volume ${\cal V}/\mu$. We now have
\begin{align}
\label{eq:Ur2}
U_{r} &\approx \sum_{k=0}^K\frac{\left(-i t {\cal V}\right)^k}{r^k\mu^{k} k!} \left(\sum_{\rho=1}^{\mu} {\cal H}\left(\vec z_\rho\right) \right)^k \nonumber \\
& = \sum_{k=0}^K\frac{\left(-i t {\cal V}\right)^k}{r^k\mu^{k} k!} \!\!\sum_{\rho_1,\cdots, \rho_k=1}^{\mu} \!\!{\cal H}\left(\vec z_{\rho_1}\right) \cdots {\cal H}\left(\vec z_{\rho_k}\right).
\end{align}
For the second quantized Hamiltonian, the $W_\gamma$ in \eq{unit_sum} are integrals over scalar functions $w_\gamma (\vec z)$ as in \eq{int_w}. Using this property it is clear that
\begin{equation}
\label{eq:calH}
{\cal H}\left(\vec z\right) = \sum_{\gamma=1}^{\Gamma} w_\gamma\left(\vec z\right) H_\gamma
\end{equation}
and the segment $U_r$ can be expressed as
\begin{align}
\label{eq:Ur3}
U_r \approx & \sum_{k=0}^K\frac{\left(-i t{\cal V}\right)^k}{r^k\mu^{k} k!} \! \! \!  \sum_{\gamma_1, \cdots, \gamma_k=1}^{\Gamma} \sum_{\rho_1, \cdots, \rho_k=1}^{\mu}\nonumber \\
& \quad \times w_{\gamma_1}\left(\vec z_{\rho_1}\right) \cdots w_{\gamma_k} \left(\vec z_{\rho_k}\right) H_{\gamma_1} \cdots H_{\gamma_k}.
\end{align}

The second quantized Hamiltonian given in \eq{3rd} is a sum of terms which are integrals over one spatial coordinate and terms which are integrals over two spatial coordinates. This case is easily accounted for by taking $\vec z$ to be a vector including both spatial coordinates, and ${\cal V}$ to be the product of the volumes for the two coordinates. One can take the terms with the integral over the single spatial coordinate to also be integrated over the second spatial coordinate, and divided by the volume of integration of that second coordinate to give the correct normalization. We may now proceed with the truncated Taylor series simulation as in \sec{simulation}. Whereas our database algorithm required $\textsc{prepare}(W)$ to create a superposition of states weighted by the $W_{\gamma}$, as in \eq{prepareW}, our on-the-fly algorithm will need to create a superposition of states weighted by the scalar integrands $w_{\gamma}(\vec z_{\rho})$.

We now show a method for dynamically decomposing each $w_\gamma(\vec z)$ into a sum of terms which differ only by a sign as in \eq{unit_decomp} so that each $w_\gamma (\vec z)$ is a sum of $M \in \Theta (\max_{\vec z,\gamma} |w_\gamma (\vec z)| / \zeta)$ terms $w_{\gamma, m} (\vec z) \in \{-1, +1\}$ where $\zeta$ is the precision of the decomposition. First, we round each entry of $w_\gamma (\vec z)$ to the nearest multiple of $2\, \zeta$ and decompose it in such a fashion that
\begin{equation}
\left | w_\gamma\left(\vec z\right) - \zeta \sum_{m = 1}^{M} w_{\gamma, m}\left(\vec z\right) \right |  \leq \zeta.
\end{equation}
To construct $\textrm{prepare}(w)$, as in \eq{preparew}, we perform logic on the output of $\textsc{sample}(w)$ (introduced in \eq{samplew}, and detailed in \sec{integrands}) which determines whether the $w_{\gamma,m}(\vec z)$ for a given value of $m$ should be $1$ or $-1$. Since the superposition in \eq{preparew} must be weighted by the square root of this coefficient, we need to prepare states that either do or do not have a phase factor $i$ so
\begin{align}
\label{eq:logic}
\ket{\ell}\ket{\widetilde{w}_\gamma\left(\vec z_\rho\right)} \mapsto \begin{cases}
\,\,\,\ket{\ell}\ket{\widetilde{w}_\gamma\left(\vec z_\rho\right)} &   \widetilde{w}_\gamma\left(\vec z_\rho\right) > \left( 2m-M\right)\zeta\\
i\ket{\ell}\ket{\widetilde{w}_\gamma\left(\vec z_\rho\right)} & \widetilde{w}_\gamma\left(\vec z_\rho\right) \leq \left(2m-M\right)\zeta
\end{cases}
\end{align}
where $\ket{\ell} = \ket{\gamma}\ket{m}\ket{\rho}$ and $\ket{\widetilde{w}_\gamma\left(\vec z_\rho\right)}$ was obtained from $\textsc{sample}(w)$. The phase factor can be obtained using phase-kickback in the usual way.
Then we apply $\textsc{sample}(w)$ a second time to erase the register $\ket{\widetilde{w}_\ell (\vec z_\rho)}$. A single query to this circuit allows for the construction of $\textsc{prepare}(w)$ with the same complexity as $\textsc{sample}(w)$.

Before explaining the integrand circuit we briefly comment on the additional resources required for the Taylor series simulation under a discretized, position-dependent, integrand Hamiltonian. As in the constant Hamiltonian case, we need one register with $\Theta(K)$ qubits to encode $\ket{k}$ and $K$ registers of $\Theta(\log \Gamma)$ qubits to encode $\ket{\gamma_1} \cdots \ket{\gamma_k}$. However, as in the explanation below \eq{preparew} we also need extra ancilla qubits to store the value of $m$, the grid point registers, as well as the value registers which are used by the integrand oracle $\textsc{sample}(w)$ in \eq{samplew}. This represents an additional ancilla overhead of $\Theta(K \log (M \mu))$.

The sources of simulation error are also similar to the constant Hamiltonian case. As we show in \sec{integrands}, we can
approximate the integrals with discrete sums to precision $\epsilon$ at a cost that is logarithmic in $1/\epsilon$.
The error due to the discrete sum is controlled by the choice of $\mu$, which we need
to select so that the resulting error per segment is less than $\epsilon/r$.  The most costly integrals (due to the size of their domain) will be the two-electron integrals in \eq{double_int} which have integrands of the form
\begin{equation}
h_{ijk\ell}\left(\vec x, \vec y\right) = \frac{\varphi^*_i \left(\vec x\right) \varphi^*_j\left(\vec y\right) \varphi_{\ell}\left(\vec x\right) \varphi_k\left(\vec y\right)} {|\vec x - \vec y|},
\label{eq:integrand}
\end{equation}
where $\vec x$ and $\vec y$ represent the three spatial degrees of freedom of two separate electrons. In \sec{integrands}, we bound the cost to the quantum algorithm of estimating the corresponding integrals.

\section{The Integrand Oracle}
\label{sec:integrands}

In \sec{integrals}, we showed how one can implement the truncated Taylor series simulation technique by replacing a superposition state having amplitudes given by integrals with a superposition state having amplitudes given by their integrands, as well as a way of decomposing those integrands. We begin this section by constructing a circuit which allows us to sample from the integrands as in \eq{samplew}. First, we will need a circuit which computes values of the $N$ spin-orbital basis functions $\varphi_1(\vec z_\rho)$ to $\varphi_N(\vec z_\rho)$ at $\vec z_\rho$, a real-valued position vector at grid point $\rho$. The action of each these oracles is
\begin{equation}
\label{eq:basis_oracle}
Q_{\varphi_j} \ket{\rho}\ket{0}^{\otimes\log M} = \ket{\rho} \ket{\widetilde{\varphi}_j\left(\vec z_\rho\right)},
\end{equation}
where $\widetilde{\varphi}_j(\vec z_\rho)$ represents the binary expansion of $\varphi_j(\vec z_\rho)$ using $\log M$ qubits. We will need $N$ different circuits of this form, one for each basis function $\varphi_1(\vec z)$ to $\varphi_N(\vec z)$. Usually, the molecular spin-orbital basis functions are represented as sums of Gaussians multiplied by polynomials \cite{Helgaker2002}. In that case, the circuit $Q_{\varphi_j}$ would be a reversible classical circuit that evaluates and sums the Gaussians associated with $\varphi_j(\vec z)$. For example, in the STO-$n$G basis set, each orbital is a linear combination of $n$ Gaussian basis functions \cite{Helgaker2002}. In general, Gaussian functions may be evaluated classically with complexity that is at most polylogarithmic in $1/\epsilon$  \cite{Zimmermann2010}. The use of Gaussians is a historical precedent used because those functions are simple to integrate on a classical computer. However, the use of a Gaussian basis is not necessarily an optimal strategy for a quantum implementation. We leave open the problem of determining an optimal representation of molecular orbital basis functions for evaluation on a quantum computer and develop a strategy based on the model chemistries used in classical treatments of electronic structure.

Next, we combine $N$ different $Q_{\varphi_j}$ circuits, one for each $\varphi_j(\vec z)$, to construct a circuit ${\cal Q}$ which allows us to apply any of the $N$ basis functions. This circuit will have depth ${\cal O}(N {\rm polylog}(Nt/\epsilon))$ and may be constructed as the block diagonal operator
\begin{equation}
{\cal Q} = \prod_{j = 1}^{N} \ket{j}\bra{j}\otimes Q_{\varphi_j}.
\end{equation}
Thus, ${\cal Q}$ is a sequence of $Q_{\varphi_j}$ circuits with the spin-orbital selection completely controlled on a register encoding the basis function index $j$.
There will a factor of $\log (Nt/\epsilon)$ because the controlled operations need to access ${\cal O}(\log N)$ qubits for $j$, as well as ${\cal O}(\log (Nt/\epsilon))$ qubits storing the position $\vec z$.
In addition, the circuit needs to perform analytic operations (e.g.\ calculating exponentials for STO-$n$G), which will contribute an additional factor polynomial in $\log (Nt/\epsilon)$.

We now discuss how one can use ${\cal Q}$ to compute the two-electron integrands in \eq{integrand}. To avoid singularities that would occur when two electrons occupy the same point in space, we change variables in \eq{integrand} so that $\vec \rad = \vec x - \vec y$. With this substitution, the integral becomes
\begin{equation}
\int\frac{\varphi_i^*(\vec x)\varphi_j^*(\vec x - \vec \rad)\varphi_\ell(\vec x)\varphi_k(\vec x - \vec \rad)}{|\vec\rad |} \,d^3 \vec \rad \, d^3 \vec x.
\end{equation}
Expressing $\vec \rad$ in spherical polar coordinates, with $\rad\equiv |\vec\rad|$, we have
\begin{align}
\int \! \varphi_i^* (\vec x)\varphi_j^*(\vec x - \vec \rad)\varphi_\ell(\vec x)\varphi_k(\vec x - \vec \rad) \,\rad\sin\theta  \, d \rad \, d\theta \, d\phi\,d^3 \vec x.
\label{eq:polar}
\end{align}
We define the maximum value of any spin-orbital function as $\varphi_{\max}$ and the maximum value of its derivative in any direction as $\varphi'_{\max}$. In addition, we truncate the integral at a finite distance $x_{\max}$. Now assume that we discretize $\vec x$ in intervals of size $\delta x$ along each degree of freedom. We can take the maximum value of $\rad$ to be $x_{\max}$, and choose $\delta \rad=\delta x$, $\delta \theta =\delta\phi = \delta x/x_{\max}$.

% Complexity.
The primary contribution to the complexity is in terms of the number of segments. The maximum value in the integrand of \eq{polar} is upper bounded by $x_{\max}\varphi_{\max}^4$. When discretizing the integral, each term in the sum is no larger than ${\cal O}(x_{\max} \varphi_{\max}^4\delta x^4 (\delta x / x_{\max})^2) = {\cal O}(\varphi_{\max}^4\delta x^6 / x_{\max})$ and there are ${\cal O}((x_{\max}/\delta x)^6)$ terms. Multiplying these together gives us the contribution of the integral to the scaling of our on-the-fly algorithm,
\begin{equation}
\label{eq:int_complex}
{\cal O}\left(\varphi_{\max}^4 x_{\max}^5\right),
\end{equation}
which corresponds to the factor of ${\cal V} \max_{\vec z,\gamma}|w_\gamma (\vec z ) |$ in \eq{lambda}.
But how do $\varphi_{\max}$ and $x_{\max}$ scale with $N$? The maximum values $\varphi_{\max}$ are predetermined by the model chemistry, and hence are independent of $N$.
Determining the appropriate value of $x_{\max}$ is a little more complicated.

Because the Hamiltonian is a sum of ${\cal O}(N^4)$ of the integrals, each integral should be approximated within error ${\cal O}(\epsilon/(N^4 t))$ to ensure that the final error is bounded by $\epsilon$.
Since the functions $\varphi_j(\vec z)$ can be chosen to decay exponentially, $x_{\max}$ can be chosen logarithmically in the allowable error $\epsilon$.
The quantum chemistry problem is always defined within a particular basis, specified as part of a model chemistry \cite{Helgaker2002}. The model chemistry prescribes how many spin-orbitals, how many basis functions, and what type of basis functions should be associated with each atom in a molecule. This includes a specification for parameters of the basis functions which impose a particular maximum value $\varphi_{\max}$, as well as a cutoff distance beyond which each $\varphi_j(\vec z)$ is negligibly small. However, the space between basis functions on different atoms must grow as the cube root of $N$, because the molecular volume will grow as ${\cal O}(N)$. This would imply that the value of $x_{\max}$ needed scales as
\begin{equation}
\label{eq:rmax}
x_{\max} \in {\cal O}\left(N^{1/3} \log\left(N t / \epsilon\right)\right).
\end{equation}

Nevertheless, each individual orbital $\varphi_j(\vec z)$ is non-negligible on a region that grows only as ${\cal O}(\log N)$ for a given model chemistry. It is therefore advantageous to modify the grid used for the integral so it only includes points where one of the associated orbitals is non-negligible. This can be performed at unit cost if the center of each spin-orbital function is provided in an additional register when querying the circuit ${\cal Q}$. As above, the region where the orbital can be regarded as non-negligible can be chosen logarithmically in $\epsilon$, to ensure that the overall error in the simulation is within error $\epsilon$.

To be more specific, the coordinates for $\vec x$ can be chosen to be centered around the center of orbital $\varphi_i$, with the components of $\vec x$ only going up to a maximum value scaling as
\begin{equation}
\label{eq:rmax2}
x_{\max} \in {\cal O}\left(\log\left(N t / \epsilon\right)\right).
\end{equation}
For $\vec \xi$, we only wish to take values such that $\varphi_j(\vec x-\vec \xi)$ are non-negligible.
Here it should be noted that the spherical polar coordinates are only advantageous if we are in a region where $\vec \xi$ is near zero, where the Cartesian coordinates would have a divergence.
In regions where $\vec \xi$ is large, the extra factor of $\xi$ for the integral in spherical polar coordinates increases the complexity.

Therefore, if the minimum value of $|\vec \xi|$ such that $\varphi_j(\vec x-\vec \xi)$ is non-negligible is ${\cal O}\left(\log\left(N t / \epsilon\right)\right)$, then 
the maximum value of $|\vec \xi|$ such that $\varphi_j(\vec x-\vec \xi)$ is non-negligible will also be ${\cal O}\left(\log\left(N t / \epsilon\right)\right)$.
Therefore we can use spherical polar coordinates, and obtain scaling as in \eq{int_complex} with $x_{\max}$ as in \eq{rmax2}.
On the other hand, if the minimum value of $|\vec \xi|$ such that $\varphi_j(\vec x-\vec \xi)$ is non-negligible is $\Omega\left(\log\left(N t / \epsilon\right)\right)$,
then we can use Cartesian coordinates, and the division by $|\vec\xi|$ can only lower the complexity.
We obtain a contribution to the complexity scaling as ${\cal O}\left(\varphi_{\max}^4 x_{\max}^3\right)$ with $x_{\max}$ as in \eq{rmax2}.
Here the power of $x_{\max}$ is $3$ rather than $5$, because we divide instead of multiplying by $|\vec\xi|$ as we did spherical polar coordinates.

% Discretization error
Next we consider the grid size needed to appropriately bound the error in the discretized integration.
The analysis in the case where Cartesian coordinates are used is relatively straightforward.
Considering a single block in $6$ dimensions with sides of length $\delta x$, the value of the integrand can only vary by the maximum derivative of the integrand times $\delta x$ (up to a constant factor).
The error for the approximation of the integral on this cube is therefore that maximum derivative times $\delta x^7$.
Then the number of these blocks in the integral is ${\cal O}((x_{\max}/\delta x)^6)$, giving an overall error scaling as $x_{\max}^6 \delta x$ times the maximum derivative of the integrand.

The maximum derivative of the integrand can be bounded in the following way.
For the derivative with respect to any component of $\vec x$, we obtain the derivative of the integrand scaling as
\begin{align}\label{mxder1}
{\cal O}\left( \frac{\varphi'_{\max} \varphi_{\max}^3}{x_{\max}}  \right),
\end{align}
where we have used the fact that we are only using Cartesian coordinates for $|\vec\xi|=\Omega(x_{\max})$.
For the derivative of the integrand with respect to any component of $\vec \xi$ in the numerator of the integrand, the same scaling is obtained.
For derivatives with respect to components of $\vec\xi$ in the denominator, the scaling is
\begin{align}\label{mxder2}
{\cal O}\left( \frac{\varphi_{\max}^4}{x_{\max}^2}  \right).
\end{align}
Overall, we therefore bound the error when discretizing in Cartesian coordinates as
\begin{align}\label{errbnd}
{\cal O}\left(\left(\varphi'_{\max}+\varphi_{\max}/x_{\max}\right)\varphi_{\max}^3 x_{\max}^5 \delta x\right).
\end{align}

The analysis for spherical polar coordinates is a little more subtle, but it is largely equivalent if we scale the angular variables.
It is convenient to define scaled angular variables
\begin{align}
\theta'\equiv  x_{\max}\theta , \quad \phi' \equiv x_{\max} \phi.
\end{align}
Then the discretization lengths for all variables are $\delta x$.
The volume of each block in the discretization is again $\delta x^6$, and there are ${\cal O}((x_{\max}/\delta x)^6)$ blocks.
The total error is again therefore the maximum derivative of the integrand multiplied by $x_{\max}^6 \delta x$.

The derivative of the integrand with respect to any component of $\vec x$ is again given by Eq.~\eqref{mxder1}.
Multiplication by $\xi$ gives a factor ${\cal O}(x_{\max})$, but the change of variables to $\theta'$ and $\phi'$ gives division by a factor of $x_{\max}^2$.
The derivative of the integrand with respect to $\xi$, $\theta'$ or $\phi'$ in any of the spin orbitals again gives a factor scaling as in Eq.~\eqref{mxder1}.
The derivative of the integrand with respect to $\xi$ or $\theta'$ in $\xi\sin(\theta'/x_{\max})$ scales as in Eq.~\eqref{mxder2}.

As a result, regardless of whether Cartesian coordinates are used or spherical polar coordinates, the error due to discretization is bounded as in \eqref{errbnd}.
Thus, to achieve error in the integral no larger than ${\cal O}(\epsilon/(N^4 t))$, we require that
\begin{equation}
\delta x \in \Theta\left(\frac \epsilon {N^4 t} \frac 1{\left(\varphi'_{\max}+\varphi_{\max}/x_{\max}\right)\varphi_{\max}^3x_{\max}^5}\right).
\end{equation}
The total number of terms in the sum then scales as
\begin{align}
&{\cal O}\left( \left(\frac{x_{\max}}{\delta x}\right)^6\right) \nn
&= \Theta\left( \left(\frac{N^4 t}{\epsilon} (\varphi'_{\max}+\varphi_{\max}/x_{\max})\varphi_{\max}^3x_{\max}^6\right)^6\right).
\end{align}
This is quite large, but because we only need to use a number of qubits that is the logarithm of this, it only contributes a logarithmic factor to the complexity.
Because the logarithm scales as ${\cal O}(\log(Nt/\epsilon))$, it contributes this factor to the complexity of $\textsc{sample}(w)$.

Given ${\cal Q}$, computing the integrand in \eq{polar} is straightforward. We need to call ${\cal Q}$ four times on registers that contain $\vec x$ and $\vec \rad$. We then multiply those outputs together with the value of $\rad \sin \theta$.
Next we consider how to construct a circuit for the one-electron integrals in \eq{single_int}. First, one constructs $N$ additional circuits similar to the ones in \eq{basis_oracle} that return $\nabla^2 \varphi_j (\vec z)$ as opposed to $\varphi_j(\vec z)$. These oracles are incorporated into a one-electron version of ${\cal Q}$ which is called along with a routine to compute the nuclear Coulomb interactions. The one-electron integrals have singularities at the positions of the nuclei. Similar to the two-electron integrals, these singularities can be avoided by using spherical polar coordinates.
Each term in the sum over the nuclei should use spherical polar coordinates centered at that nucleus. Selection between the one-electron and two-electron routines is specified by $\ket{\gamma}$. Putting this together, we can implement $\textsc{sample}(w)$ as in \eq{samplew} with ${\cal O}(N \log N)$ gates, and, as discussed in \sec{integrals}, $\textsc{prepare}(w)$ has the same complexity.

While the $\wcomp$ gate count of $\textsc{prepare}(w)$ is much less than the ${\cal O}(N^4)$ gate count of $\textsc{prepare}(W)$, our on-the-fly algorithm requires more segments than the database algorithm. Whereas our database algorithm requires $r = \Lambda t / \ln(2)$ segments where $\Lambda$ is the normalization in \eq{prepareW}, our on-the-fly algorithm requires $r = \lambda t / \ln(2)$ segments where $\lambda \in \Theta (\Gamma \varphi_{\max}^4 x_{\max}^5)$ is the normalization in \eq{preparew}, which is accounted for in \eq{lambda} and \eq{int_complex}. Thus, by performing the algorithm in \sec{simulation} using $\textsc{prepare}(w)$ instead of $\textsc{prepare}(W)$ and taking $r = \lambda t / \ln(2)$, we see that our on-the-fly algorithm scales as
\begin{equation}
\widetilde{\cal O}\left(r N K\right) = \widetilde{\cal O}\left( N K \lambda t\right).
\end{equation}
Using the scaling in \eq{rmax2}, we can bound $\lambda$ as
\begin{equation}
\label{eq:lambda2}
\lambda \in {\cal O}\left(\Gamma \varphi_{\max}^4 x_{\max}^5\right) \in {\cal O}\left(N^4 [\log \left(Nt / \epsilon\right)]^5\right).
\end{equation}
so that the overall gate count of the on-the-fly algorithm scales as
\begin{equation}
\label{eq:cost}
\widetilde{\cal O}\left(N^5 K t \right) = \widetilde{\cal O}\left(N^5 t\right).
\end{equation}
Recall that the $\widetilde{\cal O}$ notation indicates that logarithmic factors have been omitted. The full scaling includes a power of the logarithm of $1/\epsilon$.

\section{Discussion}
\label{sec:conclusion}

We have introduced two novel algorithms for the simulation of molecular systems based primarily on the results of \cite{Berry2015}. Our database algorithm involves using a database to store the molecular integrals; its gate count scales as $\widetilde{\cal O}(N^8 t)$. Our on-the-fly algorithm involves computing those integrals on-the-fly; its gate count scales as $\widetilde{\cal O}(N^{5} t)$. Both represent an exponential improvement in precision over Trotter-based methods which scale as  $\widetilde{\cal O}(N^9 \sqrt{t^3 / \epsilon})$ when using reasonably low-order decompositions, and over all other approaches to date.

Specifically, our database algorithm scales like $\widetilde{\cal O}(N^4 \Lambda t)$ where we have used the bound $\Lambda \in {\cal O}(N^4)$. However, we believe this bound is very loose. As discussed in \cite{Helgaker2002,McClean2014}, the use of local basis sets leads to a number of two-electron integrals that scales as $\widetilde{\cal O}(N^2)$ in certain limits of large molecules. Accordingly, the true scaling of the database algorithm is likely to be closer to $\widetilde{\cal O}(N^6 t)$. It also seems possible that our integration scheme is suboptimal;  it is possible that it can be improved by taking account of smaller values of $h_{ijk\ell}$.

Our asymptotic analysis suggests that these algorithms will allow for the quantum simulation of molecular systems larger than would be possible using Trotter-based methods. However, numerical simulations will be crucial in order to further optimize these algorithms and better understand their scaling properties. Just as recent work showed significantly more efficient implementations of the original Trotterized quantum chemistry algorithm \cite{Wecker2014,Hastings2015,Poulin2014,McClean2014,BabbushTrotter}, we believe the implementations discussed here are far from optimal. Furthermore, just as was observed for Trotterized quantum chemistry \cite{BabbushTrotter,Poulin2014}, we believe our simulations might scale much better when only trying to simulate ground states of real molecules. In light of this, numerical simulations may indicate that the scaling for real molecules is much less than our bounds predict.

\section*{Acknowledgements}

The authors thank Jhonathan Romero Fontalvo, Jarrod McClean, David Poulin, and Nathan Wiebe for helpful comments. D.W.B.\ is funded by an Australian Research Council Future Fellowship (FT100100761). P.J.L.\ acknowledges the support of the National Science Foundation under grant number PHY-0955518. A.A.G.\ and P.J.L.\ acknowledge the support of the Air Force Office of Scientific Research under award number FA9550-12-1-0046.

\bibliographystyle{apsrev4-1}
\bibliography{library}

\begin{thebibliography}{45}%
\makeatletter
\providecommand \@ifxundefined [1]{%
 \@ifx{#1\undefined}
}%
\providecommand \@ifnum [1]{%
 \ifnum #1\expandafter \@firstoftwo
 \else \expandafter \@secondoftwo
 \fi
}%
\providecommand \@ifx [1]{%
 \ifx #1\expandafter \@firstoftwo
 \else \expandafter \@secondoftwo
 \fi
}%
\providecommand \natexlab [1]{#1}%
\providecommand \enquote  [1]{``#1''}%
\providecommand \bibnamefont  [1]{#1}%
\providecommand \bibfnamefont [1]{#1}%
\providecommand \citenamefont [1]{#1}%
\providecommand \href@noop [0]{\@secondoftwo}%
\providecommand \href [0]{\begingroup \@sanitize@url \@href}%
\providecommand \@href[1]{\@@startlink{#1}\@@href}%
\providecommand \@@href[1]{\endgroup#1\@@endlink}%
\providecommand \@sanitize@url [0]{\catcode `\\12\catcode `\$12\catcode
  `\&12\catcode `\#12\catcode `\^12\catcode `\_12\catcode `\%12\relax}%
\providecommand \@@startlink[1]{}%
\providecommand \@@endlink[0]{}%
\providecommand \url  [0]{\begingroup\@sanitize@url \@url }%
\providecommand \@url [1]{\endgroup\@href {#1}{\urlprefix }}%
\providecommand \urlprefix  [0]{URL }%
\providecommand \Eprint [0]{\href }%
\providecommand \doibase [0]{http://dx.doi.org/}%
\providecommand \selectlanguage [0]{\@gobble}%
\providecommand \bibinfo  [0]{\@secondoftwo}%
\providecommand \bibfield  [0]{\@secondoftwo}%
\providecommand \translation [1]{[#1]}%
\providecommand \BibitemOpen [0]{}%
\providecommand \bibitemStop [0]{}%
\providecommand \bibitemNoStop [0]{.\EOS\space}%
\providecommand \EOS [0]{\spacefactor3000\relax}%
\providecommand \BibitemShut  [1]{\csname bibitem#1\endcsname}%
\let\auto@bib@innerbib\@empty
%</preamble>
\bibitem [{\citenamefont {Barends}\ \emph {et~al.}(2014)\citenamefont
  {Barends}, \citenamefont {Kelly}, \citenamefont {Megrant}, \citenamefont
  {Veitia}, \citenamefont {Sank}, \citenamefont {Jeffrey}, \citenamefont
  {White}, \citenamefont {Mutus}, \citenamefont {Fowler}, \citenamefont
  {{Campbell Chen}}, \citenamefont {Chen}, \citenamefont {Chiaro},
  \citenamefont {Dunsworth}, \citenamefont {Neill}, \citenamefont {O'Malley},
  \citenamefont {Roushan}, \citenamefont {Vainsencher}, \citenamefont {Wenner},
  \citenamefont {Korotkov}, \citenamefont {Cleland},\ and\ \citenamefont
  {Martinis}}]{Martinis2014}%
  \BibitemOpen
  \bibfield  {author} {\bibinfo {author} {\bibfnamefont {R.}~\bibnamefont
  {Barends}}, \bibinfo {author} {\bibfnamefont {J.}~\bibnamefont {Kelly}},
  \bibinfo {author} {\bibfnamefont {A.}~\bibnamefont {Megrant}}, \bibinfo
  {author} {\bibfnamefont {A.}~\bibnamefont {Veitia}}, \bibinfo {author}
  {\bibfnamefont {D.}~\bibnamefont {Sank}}, \bibinfo {author} {\bibfnamefont
  {E.}~\bibnamefont {Jeffrey}}, \bibinfo {author} {\bibfnamefont {T.~C.}\
  \bibnamefont {White}}, \bibinfo {author} {\bibfnamefont {J.}~\bibnamefont
  {Mutus}}, \bibinfo {author} {\bibfnamefont {A.~G.}\ \bibnamefont {Fowler}},
  \bibinfo {author} {\bibfnamefont {Y.}~\bibnamefont {{Campbell Chen}}},
  \bibinfo {author} {\bibfnamefont {Z.}~\bibnamefont {Chen}}, \bibinfo {author}
  {\bibfnamefont {B.}~\bibnamefont {Chiaro}}, \bibinfo {author} {\bibfnamefont
  {A.}~\bibnamefont {Dunsworth}}, \bibinfo {author} {\bibfnamefont
  {C.}~\bibnamefont {Neill}}, \bibinfo {author} {\bibfnamefont
  {P.}~\bibnamefont {O'Malley}}, \bibinfo {author} {\bibfnamefont
  {P.}~\bibnamefont {Roushan}}, \bibinfo {author} {\bibfnamefont
  {A.}~\bibnamefont {Vainsencher}}, \bibinfo {author} {\bibfnamefont
  {J.}~\bibnamefont {Wenner}}, \bibinfo {author} {\bibfnamefont {A.~N.}\
  \bibnamefont {Korotkov}}, \bibinfo {author} {\bibfnamefont {A.~N.}\
  \bibnamefont {Cleland}}, \ and\ \bibinfo {author} {\bibfnamefont
  {J.}~\bibnamefont {Martinis}},\ }\href {\doibase 10.1038/nature13171}
  {\bibfield  {journal} {\bibinfo  {journal} {Nature}\ }\textbf {\bibinfo
  {volume} {508}},\ \bibinfo {pages} {500} (\bibinfo {year}
  {2014})}\BibitemShut {NoStop}%
\bibitem [{\citenamefont {Kelly}\ \emph {et~al.}(2015)\citenamefont {Kelly},
  \citenamefont {Barends}, \citenamefont {Fowler}, \citenamefont {Megrant},
  \citenamefont {Jeffrey}, \citenamefont {White}, \citenamefont {Sank},
  \citenamefont {Mutus}, \citenamefont {Campbell}, \citenamefont {Chen},
  \citenamefont {Chen}, \citenamefont {Chiaro}, \citenamefont {Dunsworth},
  \citenamefont {Hoi}, \citenamefont {Neill}, \citenamefont {O'Malley},
  \citenamefont {Quintana}, \citenamefont {Roushan}, \citenamefont
  {Vainsencher}, \citenamefont {Wenner}, \citenamefont {Cleland},\ and\
  \citenamefont {Martinis}}]{Martinis2015}%
  \BibitemOpen
  \bibfield  {author} {\bibinfo {author} {\bibfnamefont {J.}~\bibnamefont
  {Kelly}}, \bibinfo {author} {\bibfnamefont {R.}~\bibnamefont {Barends}},
  \bibinfo {author} {\bibfnamefont {A.~G.}\ \bibnamefont {Fowler}}, \bibinfo
  {author} {\bibfnamefont {A.}~\bibnamefont {Megrant}}, \bibinfo {author}
  {\bibfnamefont {E.}~\bibnamefont {Jeffrey}}, \bibinfo {author} {\bibfnamefont
  {T.~C.}\ \bibnamefont {White}}, \bibinfo {author} {\bibfnamefont
  {D.}~\bibnamefont {Sank}}, \bibinfo {author} {\bibfnamefont {J.~Y.}\
  \bibnamefont {Mutus}}, \bibinfo {author} {\bibfnamefont {B.}~\bibnamefont
  {Campbell}}, \bibinfo {author} {\bibfnamefont {Y.}~\bibnamefont {Chen}},
  \bibinfo {author} {\bibfnamefont {Z.}~\bibnamefont {Chen}}, \bibinfo {author}
  {\bibfnamefont {B.}~\bibnamefont {Chiaro}}, \bibinfo {author} {\bibfnamefont
  {A.}~\bibnamefont {Dunsworth}}, \bibinfo {author} {\bibfnamefont {I.-C.}\
  \bibnamefont {Hoi}}, \bibinfo {author} {\bibfnamefont {C.}~\bibnamefont
  {Neill}}, \bibinfo {author} {\bibfnamefont {P.~J.~J.}\ \bibnamefont
  {O'Malley}}, \bibinfo {author} {\bibfnamefont {C.}~\bibnamefont {Quintana}},
  \bibinfo {author} {\bibfnamefont {P.}~\bibnamefont {Roushan}}, \bibinfo
  {author} {\bibfnamefont {A.}~\bibnamefont {Vainsencher}}, \bibinfo {author}
  {\bibfnamefont {J.}~\bibnamefont {Wenner}}, \bibinfo {author} {\bibfnamefont
  {A.~N.}\ \bibnamefont {Cleland}}, \ and\ \bibinfo {author} {\bibfnamefont
  {J.~M.}\ \bibnamefont {Martinis}},\ }\href {\doibase 10.1038/nature14270}
  {\bibfield  {journal} {\bibinfo  {journal} {Nature}\ }\textbf {\bibinfo
  {volume} {519}},\ \bibinfo {pages} {66} (\bibinfo {year} {2015})}\BibitemShut
  {NoStop}%
\bibitem [{\citenamefont {Nigg}\ \emph {et~al.}(2014)\citenamefont {Nigg},
  \citenamefont {M\"{u}ller}, \citenamefont {Martinez}, \citenamefont
  {Schindler}, \citenamefont {Hennrich}, \citenamefont {Monz}, \citenamefont
  {Martin-Delgado},\ and\ \citenamefont {Blatt}}]{Nigg2014}%
  \BibitemOpen
  \bibfield  {author} {\bibinfo {author} {\bibfnamefont {D.}~\bibnamefont
  {Nigg}}, \bibinfo {author} {\bibfnamefont {M.}~\bibnamefont {M\"{u}ller}},
  \bibinfo {author} {\bibfnamefont {E.~A.}\ \bibnamefont {Martinez}}, \bibinfo
  {author} {\bibfnamefont {P.}~\bibnamefont {Schindler}}, \bibinfo {author}
  {\bibfnamefont {M.}~\bibnamefont {Hennrich}}, \bibinfo {author}
  {\bibfnamefont {T.}~\bibnamefont {Monz}}, \bibinfo {author} {\bibfnamefont
  {M.~A.}\ \bibnamefont {Martin-Delgado}}, \ and\ \bibinfo {author}
  {\bibfnamefont {R.}~\bibnamefont {Blatt}},\ }\href {\doibase
  10.1126/science.1253742} {\bibfield  {journal} {\bibinfo  {journal}
  {Science}\ }\textbf {\bibinfo {volume} {345}},\ \bibinfo {pages} {302}
  (\bibinfo {year} {2014})}\BibitemShut {NoStop}%
\bibitem [{\citenamefont {C\'{o}rcoles}\ \emph {et~al.}(2015)\citenamefont
  {C\'{o}rcoles}, \citenamefont {Magesan}, \citenamefont {Srinivasan},
  \citenamefont {Cross}, \citenamefont {Steffen}, \citenamefont {Gambetta},\
  and\ \citenamefont {Chow}}]{Corcoles2015}%
  \BibitemOpen
  \bibfield  {author} {\bibinfo {author} {\bibfnamefont {A.}~\bibnamefont
  {C\'{o}rcoles}}, \bibinfo {author} {\bibfnamefont {E.}~\bibnamefont
  {Magesan}}, \bibinfo {author} {\bibfnamefont {S.~J.}\ \bibnamefont
  {Srinivasan}}, \bibinfo {author} {\bibfnamefont {A.~W.}\ \bibnamefont
  {Cross}}, \bibinfo {author} {\bibfnamefont {M.}~\bibnamefont {Steffen}},
  \bibinfo {author} {\bibfnamefont {J.~M.}\ \bibnamefont {Gambetta}}, \ and\
  \bibinfo {author} {\bibfnamefont {J.~M.}\ \bibnamefont {Chow}},\ }\href
  {\doibase 10.1038/ncomms7979} {\bibfield  {journal} {\bibinfo  {journal}
  {Nat. Commun.}\ }\textbf {\bibinfo {volume} {6}},\ \bibinfo {pages} {6979}
  (\bibinfo {year} {2015})}\BibitemShut {NoStop}%
\bibitem [{\citenamefont {Wecker}\ \emph {et~al.}(2014)\citenamefont {Wecker},
  \citenamefont {Bauer}, \citenamefont {Clark}, \citenamefont {Hastings},\ and\
  \citenamefont {Troyer}}]{Wecker2014}%
  \BibitemOpen
  \bibfield  {author} {\bibinfo {author} {\bibfnamefont {D.}~\bibnamefont
  {Wecker}}, \bibinfo {author} {\bibfnamefont {B.}~\bibnamefont {Bauer}},
  \bibinfo {author} {\bibfnamefont {B.~K.}\ \bibnamefont {Clark}}, \bibinfo
  {author} {\bibfnamefont {M.~B.}\ \bibnamefont {Hastings}}, \ and\ \bibinfo
  {author} {\bibfnamefont {M.}~\bibnamefont {Troyer}},\ }\href {\doibase
  10.1103/PhysRevA.90.022305} {\bibfield  {journal} {\bibinfo  {journal} {Phys.
  Rev. A}\ }\textbf {\bibinfo {volume} {90}},\ \bibinfo {pages} {022305}
  (\bibinfo {year} {2014})}\BibitemShut {NoStop}%
\bibitem [{\citenamefont {Hastings}\ \emph {et~al.}(2015)\citenamefont
  {Hastings}, \citenamefont {Wecker}, \citenamefont {Bauer},\ and\
  \citenamefont {Troyer}}]{Hastings2015}%
  \BibitemOpen
  \bibfield  {author} {\bibinfo {author} {\bibfnamefont {M.~B.}\ \bibnamefont
  {Hastings}}, \bibinfo {author} {\bibfnamefont {D.}~\bibnamefont {Wecker}},
  \bibinfo {author} {\bibfnamefont {B.}~\bibnamefont {Bauer}}, \ and\ \bibinfo
  {author} {\bibfnamefont {M.}~\bibnamefont {Troyer}},\ }\href
  {http://www.rintonpress.com/xxqic15/qic-15-12/0001-0021.pdf} {\bibfield
  {journal} {\bibinfo  {journal} {Quantum Inf. Comput.}\ }\textbf {\bibinfo
  {volume} {15}},\ \bibinfo {pages} {1} (\bibinfo {year} {2015})}\BibitemShut
  {NoStop}%
\bibitem [{\citenamefont {Poulin}\ \emph {et~al.}(2015)\citenamefont {Poulin},
  \citenamefont {Hastings}, \citenamefont {Wecker}, \citenamefont {Wiebe},
  \citenamefont {Doherty},\ and\ \citenamefont {Troyer}}]{Poulin2014}%
  \BibitemOpen
  \bibfield  {author} {\bibinfo {author} {\bibfnamefont {D.}~\bibnamefont
  {Poulin}}, \bibinfo {author} {\bibfnamefont {M.~B.}\ \bibnamefont
  {Hastings}}, \bibinfo {author} {\bibfnamefont {D.}~\bibnamefont {Wecker}},
  \bibinfo {author} {\bibfnamefont {N.}~\bibnamefont {Wiebe}}, \bibinfo
  {author} {\bibfnamefont {A.~C.}\ \bibnamefont {Doherty}}, \ and\ \bibinfo
  {author} {\bibfnamefont {M.}~\bibnamefont {Troyer}},\ }\href
  {http://www.rintonpress.com/xxqic15/qic-15-56/0361-0384.pdf} {\bibfield
  {journal} {\bibinfo  {journal} {Quantum Inf. Comput.}\ }\textbf {\bibinfo
  {volume} {15}},\ \bibinfo {pages} {361} (\bibinfo {year} {2015})}\BibitemShut
  {NoStop}%
\bibitem [{\citenamefont {McClean}\ \emph {et~al.}(2014)\citenamefont
  {McClean}, \citenamefont {Babbush}, \citenamefont {Love},\ and\ \citenamefont
  {Aspuru-Guzik}}]{McClean2014}%
  \BibitemOpen
  \bibfield  {author} {\bibinfo {author} {\bibfnamefont {J.~R.}\ \bibnamefont
  {McClean}}, \bibinfo {author} {\bibfnamefont {R.}~\bibnamefont {Babbush}},
  \bibinfo {author} {\bibfnamefont {P.~J.}\ \bibnamefont {Love}}, \ and\
  \bibinfo {author} {\bibfnamefont {A.}~\bibnamefont {Aspuru-Guzik}},\ }\href
  {\doibase 10.1021/jz501649m} {\bibfield  {journal} {\bibinfo  {journal} {J.
  Phys. Chem. Lett.}\ }\textbf {\bibinfo {volume} {5}},\ \bibinfo {pages}
  {4368} (\bibinfo {year} {2014})}\BibitemShut {NoStop}%
\bibitem [{\citenamefont {Babbush}\ \emph
  {et~al.}(2015{\natexlab{a}})\citenamefont {Babbush}, \citenamefont {McClean},
  \citenamefont {Wecker}, \citenamefont {Aspuru-Guzik},\ and\ \citenamefont
  {Wiebe}}]{BabbushTrotter}%
  \BibitemOpen
  \bibfield  {author} {\bibinfo {author} {\bibfnamefont {R.}~\bibnamefont
  {Babbush}}, \bibinfo {author} {\bibfnamefont {J.}~\bibnamefont {McClean}},
  \bibinfo {author} {\bibfnamefont {D.}~\bibnamefont {Wecker}}, \bibinfo
  {author} {\bibfnamefont {A.}~\bibnamefont {Aspuru-Guzik}}, \ and\ \bibinfo
  {author} {\bibfnamefont {N.}~\bibnamefont {Wiebe}},\ }\href {\doibase
  10.1103/PhysRevA.91.022311} {\bibfield  {journal} {\bibinfo  {journal} {Phys.
  Rev. A}\ }\textbf {\bibinfo {volume} {91}},\ \bibinfo {pages} {022311}
  (\bibinfo {year} {2015}{\natexlab{a}})}\BibitemShut {NoStop}%
\bibitem [{\citenamefont {Berry}\ \emph
  {et~al.}(2007{\natexlab{a}})\citenamefont {Berry}, \citenamefont {Ahokas},
  \citenamefont {Cleve},\ and\ \citenamefont {Sanders}}]{Berry2006}%
  \BibitemOpen
  \bibfield  {author} {\bibinfo {author} {\bibfnamefont {D.}~\bibnamefont
  {Berry}}, \bibinfo {author} {\bibfnamefont {G.}~\bibnamefont {Ahokas}},
  \bibinfo {author} {\bibfnamefont {R.}~\bibnamefont {Cleve}}, \ and\ \bibinfo
  {author} {\bibfnamefont {B.}~\bibnamefont {Sanders}},\ }\href {\doibase
  10.1007/s00220-006-0150-x} {\bibfield  {journal} {\bibinfo  {journal}
  {Communications in Mathematical Physics}\ }\textbf {\bibinfo {volume}
  {270}},\ \bibinfo {pages} {359} (\bibinfo {year}
  {2007}{\natexlab{a}})}\BibitemShut {NoStop}%
\bibitem [{\citenamefont {Wiebe}\ \emph {et~al.}(2011)\citenamefont {Wiebe},
  \citenamefont {Berry}, \citenamefont {Hoyer},\ and\ \citenamefont
  {Sanders}}]{Wiebe2011}%
  \BibitemOpen
  \bibfield  {author} {\bibinfo {author} {\bibfnamefont {N.}~\bibnamefont
  {Wiebe}}, \bibinfo {author} {\bibfnamefont {D.~W.}\ \bibnamefont {Berry}},
  \bibinfo {author} {\bibfnamefont {P.}~\bibnamefont {Hoyer}}, \ and\ \bibinfo
  {author} {\bibfnamefont {B.~C.}\ \bibnamefont {Sanders}},\ }\href {\doibase
  10.1088/1751-8113/44/44/445308} {\bibfield  {journal} {\bibinfo  {journal}
  {J. Phys. A Math. Theor.}\ }\textbf {\bibinfo {volume} {44}},\ \bibinfo
  {pages} {445308} (\bibinfo {year} {2011})}\BibitemShut {NoStop}%
\bibitem [{\citenamefont {Gibney}(2014)}]{Gibney2014}%
  \BibitemOpen
  \bibfield  {author} {\bibinfo {author} {\bibfnamefont {E.}~\bibnamefont
  {Gibney}},\ }\href {\doibase 10.1038/516024a} {\bibfield  {journal} {\bibinfo
   {journal} {Nature}\ }\textbf {\bibinfo {volume} {516}},\ \bibinfo {pages}
  {24} (\bibinfo {year} {2014})}\BibitemShut {NoStop}%
\bibitem [{\citenamefont {Mueck}(2015)}]{Mueck2015}%
  \BibitemOpen
  \bibfield  {author} {\bibinfo {author} {\bibfnamefont {L.}~\bibnamefont
  {Mueck}},\ }\href {\doibase 10.1038/nchem.2248} {\bibfield  {journal}
  {\bibinfo  {journal} {Nat. Chem.}\ }\textbf {\bibinfo {volume} {7}},\
  \bibinfo {pages} {361} (\bibinfo {year} {2015})}\BibitemShut {NoStop}%
\bibitem [{\citenamefont {Lloyd}(1996)}]{Lloyd1996}%
  \BibitemOpen
  \bibfield  {author} {\bibinfo {author} {\bibfnamefont {S.}~\bibnamefont
  {Lloyd}},\ }\href {\doibase 10.1126/science.273.5278.1073} {\bibfield
  {journal} {\bibinfo  {journal} {Science}\ }\textbf {\bibinfo {volume}
  {273}},\ \bibinfo {pages} {1073} (\bibinfo {year} {1996})}\BibitemShut
  {NoStop}%
\bibitem [{\citenamefont {Abrams}\ and\ \citenamefont
  {Lloyd}(1997)}]{Abrams1997}%
  \BibitemOpen
  \bibfield  {author} {\bibinfo {author} {\bibfnamefont {D.~S.}\ \bibnamefont
  {Abrams}}\ and\ \bibinfo {author} {\bibfnamefont {S.}~\bibnamefont {Lloyd}},\
  }\href {\doibase 10.1103/PhysRevLett.79.2586} {\bibfield  {journal} {\bibinfo
   {journal} {Phys. Rev. Lett.}\ }\textbf {\bibinfo {volume} {79}},\ \bibinfo
  {pages} {2586} (\bibinfo {year} {1997})}\BibitemShut {NoStop}%
\bibitem [{\citenamefont {Aspuru-Guzik}\ \emph {et~al.}(2005)\citenamefont
  {Aspuru-Guzik}, \citenamefont {Dutoi}, \citenamefont {Love},\ and\
  \citenamefont {Head-Gordon}}]{Aspuru-Guzik2005}%
  \BibitemOpen
  \bibfield  {author} {\bibinfo {author} {\bibfnamefont {A.}~\bibnamefont
  {Aspuru-Guzik}}, \bibinfo {author} {\bibfnamefont {A.~D.}\ \bibnamefont
  {Dutoi}}, \bibinfo {author} {\bibfnamefont {P.~J.}\ \bibnamefont {Love}}, \
  and\ \bibinfo {author} {\bibfnamefont {M.}~\bibnamefont {Head-Gordon}},\
  }\href {\doibase 10.1126/science.1113479} {\bibfield  {journal} {\bibinfo
  {journal} {Science}\ }\textbf {\bibinfo {volume} {309}},\ \bibinfo {pages}
  {1704} (\bibinfo {year} {2005})}\BibitemShut {NoStop}%
\bibitem [{\citenamefont {Whitfield}\ \emph {et~al.}(2011)\citenamefont
  {Whitfield}, \citenamefont {Biamonte},\ and\ \citenamefont
  {Aspuru-Guzik}}]{Whitfield2010}%
  \BibitemOpen
  \bibfield  {author} {\bibinfo {author} {\bibfnamefont {J.~D.}\ \bibnamefont
  {Whitfield}}, \bibinfo {author} {\bibfnamefont {J.}~\bibnamefont {Biamonte}},
  \ and\ \bibinfo {author} {\bibfnamefont {A.}~\bibnamefont {Aspuru-Guzik}},\
  }\href {\doibase 10.1080/00268976.2011.552441} {\bibfield  {journal}
  {\bibinfo  {journal} {Mol. Phys.}\ }\textbf {\bibinfo {volume} {109}},\
  \bibinfo {pages} {735} (\bibinfo {year} {2011})}\BibitemShut {NoStop}%
\bibitem [{\citenamefont {Babbush}\ \emph {et~al.}(2014)\citenamefont
  {Babbush}, \citenamefont {Love},\ and\ \citenamefont
  {Aspuru-Guzik}}]{BabbushAQChem}%
  \BibitemOpen
  \bibfield  {author} {\bibinfo {author} {\bibfnamefont {R.}~\bibnamefont
  {Babbush}}, \bibinfo {author} {\bibfnamefont {P.~J.}\ \bibnamefont {Love}}, \
  and\ \bibinfo {author} {\bibfnamefont {A.}~\bibnamefont {Aspuru-Guzik}},\
  }\href {\doibase 10.1038/srep06603} {\bibfield  {journal} {\bibinfo
  {journal} {Sci. Rep.}\ }\textbf {\bibinfo {volume} {4}},\ \bibinfo {pages}
  {6603} (\bibinfo {year} {2014})}\BibitemShut {NoStop}%
\bibitem [{\citenamefont {Peruzzo}\ \emph {et~al.}(2014)\citenamefont
  {Peruzzo}, \citenamefont {McClean}, \citenamefont {Shadbolt}, \citenamefont
  {Yung}, \citenamefont {Zhou}, \citenamefont {Love}, \citenamefont
  {Aspuru-Guzik},\ and\ \citenamefont {O’Brien}}]{Peruzzo2013}%
  \BibitemOpen
  \bibfield  {author} {\bibinfo {author} {\bibfnamefont {A.}~\bibnamefont
  {Peruzzo}}, \bibinfo {author} {\bibfnamefont {J.}~\bibnamefont {McClean}},
  \bibinfo {author} {\bibfnamefont {P.}~\bibnamefont {Shadbolt}}, \bibinfo
  {author} {\bibfnamefont {M.-H.}\ \bibnamefont {Yung}}, \bibinfo {author}
  {\bibfnamefont {X.-Q.}\ \bibnamefont {Zhou}}, \bibinfo {author}
  {\bibfnamefont {P.~J.}\ \bibnamefont {Love}}, \bibinfo {author}
  {\bibfnamefont {A.}~\bibnamefont {Aspuru-Guzik}}, \ and\ \bibinfo {author}
  {\bibfnamefont {J.~L.}\ \bibnamefont {O’Brien}},\ }\href {\doibase
  10.1038/ncomms5213} {\bibfield  {journal} {\bibinfo  {journal} {Nat.
  Commun.}\ }\textbf {\bibinfo {volume} {5}},\ \bibinfo {pages} {4213}
  (\bibinfo {year} {2014})}\BibitemShut {NoStop}%
\bibitem [{\citenamefont {{Cody Jones}}\ \emph {et~al.}(2012)\citenamefont
  {{Cody Jones}}, \citenamefont {Whitfield}, \citenamefont {McMahon},
  \citenamefont {Yung}, \citenamefont {Meter}, \citenamefont {Aspuru-Guzik},\
  and\ \citenamefont {Yamamoto}}]{Jones2012}%
  \BibitemOpen
  \bibfield  {author} {\bibinfo {author} {\bibfnamefont {N.}~\bibnamefont
  {{Cody Jones}}}, \bibinfo {author} {\bibfnamefont {J.~D.}\ \bibnamefont
  {Whitfield}}, \bibinfo {author} {\bibfnamefont {P.~L.}\ \bibnamefont
  {McMahon}}, \bibinfo {author} {\bibfnamefont {M.-H.}\ \bibnamefont {Yung}},
  \bibinfo {author} {\bibfnamefont {R.~V.}\ \bibnamefont {Meter}}, \bibinfo
  {author} {\bibfnamefont {A.}~\bibnamefont {Aspuru-Guzik}}, \ and\ \bibinfo
  {author} {\bibfnamefont {Y.}~\bibnamefont {Yamamoto}},\ }\href {\doibase
  10.1088/1367-2630/14/11/115023} {\bibfield  {journal} {\bibinfo  {journal}
  {New J. Phys.}\ }\textbf {\bibinfo {volume} {14}},\ \bibinfo {pages} {115023}
  (\bibinfo {year} {2012})}\BibitemShut {NoStop}%
\bibitem [{\citenamefont {Veis}\ and\ \citenamefont
  {Pittner}(2010)}]{Veis2010}%
  \BibitemOpen
  \bibfield  {author} {\bibinfo {author} {\bibfnamefont {L.}~\bibnamefont
  {Veis}}\ and\ \bibinfo {author} {\bibfnamefont {J.}~\bibnamefont {Pittner}},\
  }\href {\doibase 10.1063/1.3503767} {\bibfield  {journal} {\bibinfo
  {journal} {J. Chem. Phys.}\ }\textbf {\bibinfo {volume} {133}},\ \bibinfo
  {pages} {194106} (\bibinfo {year} {2010})}\BibitemShut {NoStop}%
\bibitem [{\citenamefont {Wang}\ \emph {et~al.}(2015)\citenamefont {Wang},
  \citenamefont {Dolde}, \citenamefont {Biamonte}, \citenamefont {Babbush},
  \citenamefont {Bergholm}, \citenamefont {Yang}, \citenamefont {Jakobi},
  \citenamefont {Neumann}, \citenamefont {Aspuru-Guzik}, \citenamefont
  {Whitfield},\ and\ \citenamefont {Wrachtrup}}]{Wang2014}%
  \BibitemOpen
  \bibfield  {author} {\bibinfo {author} {\bibfnamefont {Y.}~\bibnamefont
  {Wang}}, \bibinfo {author} {\bibfnamefont {F.}~\bibnamefont {Dolde}},
  \bibinfo {author} {\bibfnamefont {J.}~\bibnamefont {Biamonte}}, \bibinfo
  {author} {\bibfnamefont {R.}~\bibnamefont {Babbush}}, \bibinfo {author}
  {\bibfnamefont {V.}~\bibnamefont {Bergholm}}, \bibinfo {author}
  {\bibfnamefont {S.}~\bibnamefont {Yang}}, \bibinfo {author} {\bibfnamefont
  {I.}~\bibnamefont {Jakobi}}, \bibinfo {author} {\bibfnamefont
  {P.}~\bibnamefont {Neumann}}, \bibinfo {author} {\bibfnamefont
  {A.}~\bibnamefont {Aspuru-Guzik}}, \bibinfo {author} {\bibfnamefont {J.~D.}\
  \bibnamefont {Whitfield}}, \ and\ \bibinfo {author} {\bibfnamefont
  {J.}~\bibnamefont {Wrachtrup}},\ }\href {\doibase 10.1021/acsnano.5b01651}
  {\bibfield  {journal} {\bibinfo  {journal} {ACS Nano}\ } (\bibinfo {year}
  {2015}),\ 10.1021/acsnano.5b01651},\ \Eprint {http://arxiv.org/abs/1405.2696}
  {arXiv:1405.2696} \BibitemShut {NoStop}%
\bibitem [{\citenamefont {Whitfield}(2013)}]{Whitfield2013b}%
  \BibitemOpen
  \bibfield  {author} {\bibinfo {author} {\bibfnamefont {J.~D.}\ \bibnamefont
  {Whitfield}},\ }\href {\doibase 10.1063/1.4812566} {\bibfield  {journal}
  {\bibinfo  {journal} {J. Chem. Phys.}\ }\textbf {\bibinfo {volume} {139}},\
  \bibinfo {pages} {021105} (\bibinfo {year} {2013})}\BibitemShut {NoStop}%
\bibitem [{\citenamefont {Whitfield}(2015)}]{Whitfield2015}%
  \BibitemOpen
  \bibfield  {author} {\bibinfo {author} {\bibfnamefont {J.~D.}\ \bibnamefont
  {Whitfield}},\ }\href {http://arxiv.org/abs/1502.03771} {\bibfield  {journal}
  {\bibinfo  {journal} {e-print arXiv:1502.03771}\ } (\bibinfo {year}
  {2015})}\BibitemShut {NoStop}%
\bibitem [{\citenamefont {Li}\ \emph {et~al.}(2011)\citenamefont {Li},
  \citenamefont {Yung}, \citenamefont {Chen}, \citenamefont {Lu}, \citenamefont
  {Whitfield}, \citenamefont {Peng}, \citenamefont {Aspuru-Guzik},\ and\
  \citenamefont {Du}}]{Li2011}%
  \BibitemOpen
  \bibfield  {author} {\bibinfo {author} {\bibfnamefont {Z.}~\bibnamefont
  {Li}}, \bibinfo {author} {\bibfnamefont {M.-H.}\ \bibnamefont {Yung}},
  \bibinfo {author} {\bibfnamefont {H.}~\bibnamefont {Chen}}, \bibinfo {author}
  {\bibfnamefont {D.}~\bibnamefont {Lu}}, \bibinfo {author} {\bibfnamefont
  {J.~D.}\ \bibnamefont {Whitfield}}, \bibinfo {author} {\bibfnamefont
  {X.}~\bibnamefont {Peng}}, \bibinfo {author} {\bibfnamefont {A.}~\bibnamefont
  {Aspuru-Guzik}}, \ and\ \bibinfo {author} {\bibfnamefont {J.}~\bibnamefont
  {Du}},\ }\href {\doibase doi:10.1038/srep00088} {\bibfield  {journal}
  {\bibinfo  {journal} {Sci. Rep.}\ }\textbf {\bibinfo {volume} {1}},\ \bibinfo
  {pages} {88} (\bibinfo {year} {2011})}\BibitemShut {NoStop}%
\bibitem [{\citenamefont {Yung}\ \emph {et~al.}(2014)\citenamefont {Yung},
  \citenamefont {Casanova}, \citenamefont {Mezzacapo}, \citenamefont {McClean},
  \citenamefont {Lamata}, \citenamefont {Aspuru-Guzik},\ and\ \citenamefont
  {Solano}}]{Yung2013}%
  \BibitemOpen
  \bibfield  {author} {\bibinfo {author} {\bibfnamefont {M.-H.}\ \bibnamefont
  {Yung}}, \bibinfo {author} {\bibfnamefont {J.}~\bibnamefont {Casanova}},
  \bibinfo {author} {\bibfnamefont {A.}~\bibnamefont {Mezzacapo}}, \bibinfo
  {author} {\bibfnamefont {J.}~\bibnamefont {McClean}}, \bibinfo {author}
  {\bibfnamefont {L.}~\bibnamefont {Lamata}}, \bibinfo {author} {\bibfnamefont
  {A.}~\bibnamefont {Aspuru-Guzik}}, \ and\ \bibinfo {author} {\bibfnamefont
  {E.}~\bibnamefont {Solano}},\ }\href {\doibase 10.1038/srep03589} {\bibfield
  {journal} {\bibinfo  {journal} {Sci. Rep.}\ }\textbf {\bibinfo {volume}
  {4}},\ \bibinfo {pages} {3589} (\bibinfo {year} {2014})},\ \Eprint
  {http://arxiv.org/abs/1307.4326} {arXiv:1307.4326} \BibitemShut {NoStop}%
\bibitem [{\citenamefont {Toloui}\ and\ \citenamefont
  {Love}(2013)}]{Toloui2013}%
  \BibitemOpen
  \bibfield  {author} {\bibinfo {author} {\bibfnamefont {B.}~\bibnamefont
  {Toloui}}\ and\ \bibinfo {author} {\bibfnamefont {P.~J.}\ \bibnamefont
  {Love}},\ }\href {http://arxiv.org/abs/1312.2579} {\bibfield  {journal}
  {\bibinfo  {journal} {e-print arXiv:1312.2579}\ } (\bibinfo {year}
  {2013})}\BibitemShut {NoStop}%
\bibitem [{\citenamefont {Aharonov}\ and\ \citenamefont
  {Ta-Shma}(2003)}]{Aharonov2003}%
  \BibitemOpen
  \bibfield  {author} {\bibinfo {author} {\bibfnamefont {D.}~\bibnamefont
  {Aharonov}}\ and\ \bibinfo {author} {\bibfnamefont {A.}~\bibnamefont
  {Ta-Shma}},\ }in\ \href {\doibase 10.1145/780542.780546} {\emph {\bibinfo
  {booktitle} {Proceedings of the Thirty-fifth Annual ACM Symposium on Theory
  of Computing}}},\ \bibinfo {series and number} {STOC '03}\ (\bibinfo
  {publisher} {ACM},\ \bibinfo {address} {New York, NY, USA},\ \bibinfo {year}
  {2003})\ pp.\ \bibinfo {pages} {20--29}\BibitemShut {NoStop}%
\bibitem [{\citenamefont {Berry}\ \emph
  {et~al.}(2007{\natexlab{b}})\citenamefont {Berry}, \citenamefont {Ahokas},
  \citenamefont {Cleve},\ and\ \citenamefont {Sanders}}]{Berry2007}%
  \BibitemOpen
  \bibfield  {author} {\bibinfo {author} {\bibfnamefont {D.~W.}\ \bibnamefont
  {Berry}}, \bibinfo {author} {\bibfnamefont {G.}~\bibnamefont {Ahokas}},
  \bibinfo {author} {\bibfnamefont {R.}~\bibnamefont {Cleve}}, \ and\ \bibinfo
  {author} {\bibfnamefont {B.~C.}\ \bibnamefont {Sanders}},\ }\href {\doibase
  10.1007/s00220-006-0150-x} {\bibfield  {journal} {\bibinfo  {journal}
  {Commun. Math. Phys.}\ }\textbf {\bibinfo {volume} {270}},\ \bibinfo {pages}
  {359} (\bibinfo {year} {2007}{\natexlab{b}})}\BibitemShut {NoStop}%
\bibitem [{\citenamefont {Cleve}\ \emph {et~al.}(2009)\citenamefont {Cleve},
  \citenamefont {Gottesman}, \citenamefont {Mosca}, \citenamefont {Somma},\
  and\ \citenamefont {Yonge-Mallo}}]{Cleve2009}%
  \BibitemOpen
  \bibfield  {author} {\bibinfo {author} {\bibfnamefont {R.}~\bibnamefont
  {Cleve}}, \bibinfo {author} {\bibfnamefont {D.}~\bibnamefont {Gottesman}},
  \bibinfo {author} {\bibfnamefont {M.}~\bibnamefont {Mosca}}, \bibinfo
  {author} {\bibfnamefont {R.~D.}\ \bibnamefont {Somma}}, \ and\ \bibinfo
  {author} {\bibfnamefont {D.}~\bibnamefont {Yonge-Mallo}},\ }in\ \href
  {\doibase 10.1145/1536414.1536471} {\emph {\bibinfo {booktitle} {Proceedings
  of the Forty-first Annual ACM Symposium on Theory of Computing}}},\ \bibinfo
  {series and number} {STOC '09}\ (\bibinfo  {publisher} {ACM},\ \bibinfo
  {address} {New York, NY, USA},\ \bibinfo {year} {2009})\ pp.\ \bibinfo
  {pages} {409--416}\BibitemShut {NoStop}%
\bibitem [{\citenamefont {Berry}\ \emph
  {et~al.}(2014{\natexlab{a}})\citenamefont {Berry}, \citenamefont {Cleve},\
  and\ \citenamefont {Gharibian}}]{Berry2014}%
  \BibitemOpen
  \bibfield  {author} {\bibinfo {author} {\bibfnamefont {D.~W.}\ \bibnamefont
  {Berry}}, \bibinfo {author} {\bibfnamefont {R.}~\bibnamefont {Cleve}}, \ and\
  \bibinfo {author} {\bibfnamefont {S.}~\bibnamefont {Gharibian}},\ }\href
  {http://www.rintonpress.com/xxqic14/qic-14-12/0001-0030.pdf} {\bibfield
  {journal} {\bibinfo  {journal} {Quantum Inf. Comput.}\ }\textbf {\bibinfo
  {volume} {14}},\ \bibinfo {pages} {1} (\bibinfo {year}
  {2014}{\natexlab{a}})}\BibitemShut {NoStop}%
\bibitem [{\citenamefont {Berry}\ \emph
  {et~al.}(2014{\natexlab{b}})\citenamefont {Berry}, \citenamefont {Childs},
  \citenamefont {Cleve}, \citenamefont {Kothari},\ and\ \citenamefont
  {Somma}}]{Berry2013}%
  \BibitemOpen
  \bibfield  {author} {\bibinfo {author} {\bibfnamefont {D.~W.}\ \bibnamefont
  {Berry}}, \bibinfo {author} {\bibfnamefont {A.~M.}\ \bibnamefont {Childs}},
  \bibinfo {author} {\bibfnamefont {R.}~\bibnamefont {Cleve}}, \bibinfo
  {author} {\bibfnamefont {R.}~\bibnamefont {Kothari}}, \ and\ \bibinfo
  {author} {\bibfnamefont {R.~D.}\ \bibnamefont {Somma}},\ }in\ \href {\doibase
  10.1145/2591796.2591854} {\emph {\bibinfo {booktitle} {Proceedings of the
  46th Annual ACM Symposium on Theory of Computing}}},\ \bibinfo {series and
  number} {STOC '14}\ (\bibinfo  {publisher} {ACM},\ \bibinfo {address} {New
  York, NY, USA},\ \bibinfo {year} {2014})\ pp.\ \bibinfo {pages}
  {283--292}\BibitemShut {NoStop}%
\bibitem [{\citenamefont {Berry}\ \emph {et~al.}(2015)\citenamefont {Berry},
  \citenamefont {Childs}, \citenamefont {Cleve}, \citenamefont {Kothari},\ and\
  \citenamefont {Somma}}]{Berry2015}%
  \BibitemOpen
  \bibfield  {author} {\bibinfo {author} {\bibfnamefont {D.~W.}\ \bibnamefont
  {Berry}}, \bibinfo {author} {\bibfnamefont {A.~M.}\ \bibnamefont {Childs}},
  \bibinfo {author} {\bibfnamefont {R.}~\bibnamefont {Cleve}}, \bibinfo
  {author} {\bibfnamefont {R.}~\bibnamefont {Kothari}}, \ and\ \bibinfo
  {author} {\bibfnamefont {R.~D.}\ \bibnamefont {Somma}},\ }\href {\doibase
  10.1103/PhysRevLett.114.090502} {\bibfield  {journal} {\bibinfo  {journal}
  {Phys. Rev. Lett.}\ }\textbf {\bibinfo {volume} {114}},\ \bibinfo {pages}
  {090502} (\bibinfo {year} {2015})}\BibitemShut {NoStop}%
\bibitem [{\citenamefont {Jordan}\ and\ \citenamefont
  {Wigner}(1928)}]{Jordan1928}%
  \BibitemOpen
  \bibfield  {author} {\bibinfo {author} {\bibfnamefont {P.}~\bibnamefont
  {Jordan}}\ and\ \bibinfo {author} {\bibfnamefont {E.}~\bibnamefont
  {Wigner}},\ }\href {\doibase 10.1007/BF01331938} {\bibfield  {journal}
  {\bibinfo  {journal} {Zeitschrift f\"{u}r Phys.}\ }\textbf {\bibinfo {volume}
  {47}},\ \bibinfo {pages} {631} (\bibinfo {year} {1928})}\BibitemShut
  {NoStop}%
\bibitem [{\citenamefont {Somma}\ \emph {et~al.}(2002)\citenamefont {Somma},
  \citenamefont {Ortiz}, \citenamefont {Gubernatis}, \citenamefont {Knill},\
  and\ \citenamefont {Laflamme}}]{Somma2002}%
  \BibitemOpen
  \bibfield  {author} {\bibinfo {author} {\bibfnamefont {R.~D.}\ \bibnamefont
  {Somma}}, \bibinfo {author} {\bibfnamefont {G.}~\bibnamefont {Ortiz}},
  \bibinfo {author} {\bibfnamefont {J.}~\bibnamefont {Gubernatis}}, \bibinfo
  {author} {\bibfnamefont {E.}~\bibnamefont {Knill}}, \ and\ \bibinfo {author}
  {\bibfnamefont {R.}~\bibnamefont {Laflamme}},\ }\href {\doibase
  10.1103/PhysRevA.65.042323} {\bibfield  {journal} {\bibinfo  {journal} {Phys.
  Rev. A}\ }\textbf {\bibinfo {volume} {65}},\ \bibinfo {pages} {17} (\bibinfo
  {year} {2002})}\BibitemShut {NoStop}%
\bibitem [{\citenamefont {Babbush}\ \emph
  {et~al.}(2015{\natexlab{b}})\citenamefont {Babbush}, \citenamefont {Berry},
  \citenamefont {Kivlichan}, \citenamefont {Wei}, \citenamefont {Love},\ and\
  \citenamefont {Aspuru-Guzik}}]{BabbushSparse2}%
  \BibitemOpen
  \bibfield  {author} {\bibinfo {author} {\bibfnamefont {R.}~\bibnamefont
  {Babbush}}, \bibinfo {author} {\bibfnamefont {D.~W.}\ \bibnamefont {Berry}},
  \bibinfo {author} {\bibfnamefont {I.~D.}\ \bibnamefont {Kivlichan}}, \bibinfo
  {author} {\bibfnamefont {A.~Y.}\ \bibnamefont {Wei}}, \bibinfo {author}
  {\bibfnamefont {P.~J.}\ \bibnamefont {Love}}, \ and\ \bibinfo {author}
  {\bibfnamefont {A.}~\bibnamefont {Aspuru-Guzik}},\ }\href
  {http://arxiv.org/abs/1506.01029} {\bibfield  {journal} {\bibinfo  {journal}
  {e-print arXiv:1506.01029}\ } (\bibinfo {year}
  {2015}{\natexlab{b}})}\BibitemShut {NoStop}%
\bibitem [{\citenamefont {Bravyi}\ and\ \citenamefont
  {Kitaev}(2002)}]{Bravyi2002}%
  \BibitemOpen
  \bibfield  {author} {\bibinfo {author} {\bibfnamefont {S.}~\bibnamefont
  {Bravyi}}\ and\ \bibinfo {author} {\bibfnamefont {A.}~\bibnamefont
  {Kitaev}},\ }\href {\doibase 10.1006/aphy.2002.6254} {\bibfield  {journal}
  {\bibinfo  {journal} {Ann. Phys. (N. Y).}\ }\textbf {\bibinfo {volume}
  {298}},\ \bibinfo {pages} {210} (\bibinfo {year} {2002})}\BibitemShut
  {NoStop}%
\bibitem [{\citenamefont {Seeley}\ \emph {et~al.}(2012)\citenamefont {Seeley},
  \citenamefont {Richard},\ and\ \citenamefont {Love}}]{Seeley2012}%
  \BibitemOpen
  \bibfield  {author} {\bibinfo {author} {\bibfnamefont {J.~T.}\ \bibnamefont
  {Seeley}}, \bibinfo {author} {\bibfnamefont {M.~J.}\ \bibnamefont {Richard}},
  \ and\ \bibinfo {author} {\bibfnamefont {P.~J.}\ \bibnamefont {Love}},\
  }\href {\doibase 10.1063/1.4768229} {\bibfield  {journal} {\bibinfo
  {journal} {J. Chem. Phys.}\ }\textbf {\bibinfo {volume} {137}},\ \bibinfo
  {pages} {224109} (\bibinfo {year} {2012})}\BibitemShut {NoStop}%
\bibitem [{\citenamefont {Tranter}\ \emph {et~al.}(2015)\citenamefont
  {Tranter}, \citenamefont {Sofia}, \citenamefont {Seeley}, \citenamefont
  {Kaicher}, \citenamefont {McClean}, \citenamefont {Babbush}, \citenamefont
  {Coveney}, \citenamefont {Mintert}, \citenamefont {Wilhelm},\ and\
  \citenamefont {Love}}]{Tranter2015}%
  \BibitemOpen
  \bibfield  {author} {\bibinfo {author} {\bibfnamefont {A.}~\bibnamefont
  {Tranter}}, \bibinfo {author} {\bibfnamefont {S.}~\bibnamefont {Sofia}},
  \bibinfo {author} {\bibfnamefont {J.}~\bibnamefont {Seeley}}, \bibinfo
  {author} {\bibfnamefont {M.}~\bibnamefont {Kaicher}}, \bibinfo {author}
  {\bibfnamefont {J.}~\bibnamefont {McClean}}, \bibinfo {author} {\bibfnamefont
  {R.}~\bibnamefont {Babbush}}, \bibinfo {author} {\bibfnamefont {P.~V.}\
  \bibnamefont {Coveney}}, \bibinfo {author} {\bibfnamefont {F.}~\bibnamefont
  {Mintert}}, \bibinfo {author} {\bibfnamefont {F.}~\bibnamefont {Wilhelm}}, \
  and\ \bibinfo {author} {\bibfnamefont {P.~J.}\ \bibnamefont {Love}},\ }\href
  {\doibase 10.1002/qua.24969} {\bibfield  {journal} {\bibinfo  {journal}
  {International Journal of Quantum Chemistry}\ } (\bibinfo {year} {2015}),\
  10.1002/qua.24969}\BibitemShut {NoStop}%
\bibitem [{\citenamefont {Abrams}\ and\ \citenamefont
  {Williams}(1999)}]{Abrams1999fast}%
  \BibitemOpen
  \bibfield  {author} {\bibinfo {author} {\bibfnamefont {D.~S.}\ \bibnamefont
  {Abrams}}\ and\ \bibinfo {author} {\bibfnamefont {C.~P.}\ \bibnamefont
  {Williams}},\ }\href {http://arxiv.org/abs/quant-ph/9908083} {\bibfield
  {journal} {\bibinfo  {journal} {e-print arXiv:quant-ph/9908083}\ } (\bibinfo
  {year} {1999})}\BibitemShut {NoStop}%
\bibitem [{\citenamefont {Grover}(2000)}]{Grover2000}%
  \BibitemOpen
  \bibfield  {author} {\bibinfo {author} {\bibfnamefont {L.~K.}\ \bibnamefont
  {Grover}},\ }\href {\doibase 10.1103/PhysRevLett.85.1334} {\bibfield
  {journal} {\bibinfo  {journal} {Phys. Rev. Lett.}\ }\textbf {\bibinfo
  {volume} {85}},\ \bibinfo {pages} {1334} (\bibinfo {year}
  {2000})}\BibitemShut {NoStop}%
\bibitem [{\citenamefont {Grover}(1996)}]{Grover1996}%
  \BibitemOpen
  \bibfield  {author} {\bibinfo {author} {\bibfnamefont {L.~K.}\ \bibnamefont
  {Grover}},\ }in\ \href {\doibase 10.1145/237814.237866} {\emph {\bibinfo
  {booktitle} {Proceedings of the Twenty-eighth Annual ACM Symposium on Theory
  of Computing}}},\ \bibinfo {series and number} {STOC '96}\ (\bibinfo
  {publisher} {ACM},\ \bibinfo {address} {New York, NY, USA},\ \bibinfo {year}
  {1996})\ pp.\ \bibinfo {pages} {212--219}\BibitemShut {NoStop}%
\bibitem [{\citenamefont {Helgaker}\ \emph {et~al.}(2002)\citenamefont
  {Helgaker}, \citenamefont {Jorgensen},\ and\ \citenamefont
  {Olsen}}]{Helgaker2002}%
  \BibitemOpen
  \bibfield  {author} {\bibinfo {author} {\bibfnamefont {T.}~\bibnamefont
  {Helgaker}}, \bibinfo {author} {\bibfnamefont {P.}~\bibnamefont {Jorgensen}},
  \ and\ \bibinfo {author} {\bibfnamefont {J.}~\bibnamefont {Olsen}},\
  }\href@noop {} {\emph {\bibinfo {title} {{Molecular Electronic Structure
  Theory}}}}\ (\bibinfo  {publisher} {Wiley},\ \bibinfo {year}
  {2002})\BibitemShut {NoStop}%
\bibitem [{\citenamefont {Shende}\ \emph {et~al.}(2006)\citenamefont {Shende},
  \citenamefont {Bullock},\ and\ \citenamefont {Markov}}]{Shende2006}%
  \BibitemOpen
  \bibfield  {author} {\bibinfo {author} {\bibfnamefont {V.}~\bibnamefont
  {Shende}}, \bibinfo {author} {\bibfnamefont {S.}~\bibnamefont {Bullock}}, \
  and\ \bibinfo {author} {\bibfnamefont {I.}~\bibnamefont {Markov}},\ }\href
  {\doibase 10.1109/TCAD.2005.855930} {\bibfield  {journal} {\bibinfo
  {journal} {IEEE Trans. Comput. Des. Integr. Circuits Syst.}\ }\textbf
  {\bibinfo {volume} {25}},\ \bibinfo {pages} {1000} (\bibinfo {year}
  {2006})}\BibitemShut {NoStop}%
\bibitem [{\citenamefont {Brent}\ and\ \citenamefont
  {Zimmermann}(2010)}]{Zimmermann2010}%
  \BibitemOpen
  \bibfield  {author} {\bibinfo {author} {\bibfnamefont {R.~P.}\ \bibnamefont
  {Brent}}\ and\ \bibinfo {author} {\bibfnamefont {P.}~\bibnamefont
  {Zimmermann}},\ }\href@noop {} {\emph {\bibinfo {title} {{Modern Computer
  Arithmetic}}}}\ (\bibinfo  {publisher} {Cambridge Univesity Press},\ \bibinfo
  {year} {2010})\ p.\ \bibinfo {pages} {236}\BibitemShut {NoStop}%
\end{thebibliography}%

\end{document}